\documentclass[journal]{IEEEtran}
\ifCLASSINFOpdf
\else
\fi
\usepackage{graphicx}
\graphicspath{{/Figures/}}
\DeclareGraphicsExtensions{.eps,.pdf,.png,.jpg}
\usepackage{subfig}
\usepackage{amsmath}
\usepackage{algorithm}
\usepackage{algorithmic}
\usepackage{cite}
\usepackage{amsthm}
\usepackage{amssymb}
\usepackage{color}
\usepackage{multicol}
\usepackage{bm}
\usepackage{tabularx}
\usepackage{booktabs}
\usepackage{balance}
\usepackage{lettrine}

\theoremstyle{definition}

\theoremstyle{remark}

\theoremstyle{proposition}

\begin{document}
%
\title{Machine-to-Machine (M2M) Communications in Software-defined and Virtualized Cellular Networks}

\author{Meng Li, \IEEEmembership{Student Member,~IEEE}, F.~Richard~Yu, \IEEEmembership{Senior Member,~IEEE}, Pengbo Si, \IEEEmembership{Senior Member,~IEEE}, \\Enchang Sun, \IEEEmembership{Senior Member,~IEEE}, Yanhua Zhang, and Haipeng Yao, \IEEEmembership{ Member,~IEEE}\\





}


\maketitle

\begin{abstract}
Machine-to-machine (M2M) communications  have attracted great attention from both academia and industry. In this paper, with recent advances in wireless network virtualization and  software-defined networking (SDN), we propose a novel framework for M2M communications in software-defined cellular networks with wireless network virtualization. In the proposed framework, according to different functions and quality of service (QoS) requirements of machine-type communication devices (MTCDs), a hypervisor enables the virtualization of the physical M2M network, which is abstracted and sliced into multiple virtual M2M networks. In addition, we develop a decision-theoretic approach to optimize the random access process of M2M communications. Furthermore, we develop a feedback and control loop to dynamically adjust the number of resource blocks (RBs) that are used in the random access phase in a virtual M2M network by the SDN controller. Extensive simulation results with different system parameters are presented to show the performance of the proposed scheme.

\end{abstract}


%
\IEEEpeerreviewmaketitle

\section{Introduction}
\lettrine[lines=2]{M}{achine-to-machine} (M2M) communications, also named as machine-type communications (MTCs), have attracted great attention in both academia and industry~\cite{CS11}. It is estimated by the wireless world research forum (WWRF) that the number of wireless devices will increase to 7 trillion to connect various networks in the future, including a large number of  machine-type communication devices (MTCDs)~\cite{KK12}. Moreover, a report suggests that the number of M2M devices by 2020 will be around 50 billion for a projected population of around 8 billion at that time~\cite{IT14}.

Unlike traditional human-to-human (H2H) communications (e.g., voice, messages or video streaming), M2M communications have two main distinct characteristics: one is the large and rapid increasing number of MTCDs in the network (e.g., smart power grids, intelligent transportation, e-health, and surveillance)~\cite{SR15}, the other is the data transmission in each time slot, which is mostly small-sized but the frequency of their making data connections is higher than traditional communication devices due to their specific roles and functions~\cite{KL14}.\

Based on these characteristics, how to support more MTCDs simultaneously connecting and accessing to the cellular network is an important and inevitable issue~\cite{IT14}. Specifically, for M2M communications in cellular networks, two problems need to be carefully handled: one is preamble collisions on physical random access channel (PRACH), the other is resource allocation in the random access (RA) process~\cite{ZK15}. An approach to reduce preambles collision probability is proposed in~\cite{KK12} based on fixed timing alignment (TA) information. The authors of~\cite{OH15} propose a concept of random access efficiency, and formulate an optimization problem to maximize the random access efficiency with the delay constraint, according to the number of random access opportunities (RAOs) and MTCDs. In~\cite{HH13}, the authors introduce several RA overload control mechanisms to avoid collisions. The authors of~\cite{JKK14} investigate a scheme that provides additional preambles by spatially partitioning a cell coverage into multiple group regions and reducing cyclic shift size in RA preambles. In~\cite{WS12}, the authors discuss an analytical model according to collision probability and the success probability of random access by adopting the concept of RAOs.\

Although some excellent works have been done on random access with M2M communications, most existing research focus on  preamble collision avoidance mechanisms. However, in practical networks, the MTCDs may fail to access the network if there is no enough radio resource allocated to the RA process~\cite{ZK15}. Furthermore, only one class of MTCDs are considered in most existing works. However, in practical networks, different MTCDs (e.g., MTCDs used for emergency services, security services, public utilities and private utilities) have different quality of service (QoS) requirements\cite{GC15}. In addition,  resource collision or congestion may occur when they access the network simultaneously. Therefore, MTCDs should be treated differently in M2M communications~\cite{CD15,AH15}.

In this paper, with recent advances in \emph{wireless network virtualization} \cite{LY15} and \emph{software-defined networking} (SDN)~\cite{KR15}, we propose a novel framework for M2M communications in software-defined cellular networks with wireless network virtualization.
Wireless network virtualization has been considered as a promising technology for next generation wireless networks~\cite{LY15}. Using network function virtualization techniques, a physical wireless network can be abstracted and sliced into multiple virtual wireless networks, so that differentiated M2M services can be provided with differentiated QoS. Wireless network virtualization provides the momentum for new emerging design principles towards software-defined wireless networks. SDN separates the control plane from the data plane, and introduces the ability of programming the network via a logically centralized software-defined controller~\cite{CM15}. As the SDN controller has a global view of the network, radio resource can be managed efficiently in response to time-varying network conditions.  In addition, the software-defined approach allows spectrum to be managed more efficiently, since the logically centralized control can be aware of the spectrum usage in the network, and allow proper spectrum mobility and effective implementation of spectrum sharing strategies~\cite{KR15,SG15,AHSWN_YSY16,LYH10}. To the best of our knowledge, wireless network virtualization and SDN have not been well studied for M2M communications. The distinctive features of this paper are as follows.


\begin{itemize}
  \item We propose a novel framework for M2M communications in software-defined cellular networks with wireless network virtualization. In the proposed framework, according to different functions and classes of MTCDs, a hypervisor enables the virtualization of the physical M2M network, which is abstracted and sliced into multiple virtual M2M networks. In addition, through the SDN controller, network resources can be dynamically adjusted and allocated amongst virtual networks according to the functions and QoS requirements of different M2M networks.
  \item In the proposed  framework, we develop a decision-theoretic approach to optimize the random access process of M2M communications. Specifically, we formulate the random access process in M2M communications as a partially observable Markov decision process (POMDP). The maximum transmission rate can be obtained by MTCDs from the information state in the POMDP, which encapsulates the history of system state and access decision.
  \item Furthermore, we develop a feedback and control loop to dynamically adjust the number of resource blocks (RBs) that are used in the random access phase in a virtual M2M network by the SDN controller. According to the gap of ratio between the obtained and the desired transmission rate, the number of RBs is dynamically adjusted and allocated through the control loop by the  SDN controller.
  \item Extensive simulations with different system parameters are conducted to show the performance of the proposed scheme. It is shown that the system performance can be improved significantly through the POMDP optimization and the proposed feedback and control loop.
\end{itemize}

The rest of this article is organized as follows. Section \ref{sec:Overview} presents an overview of software-defined cellular networks with M2M communications and wireless network virtualization. System model is presented in Section \ref{sec:Systemmodel}. In Section \ref{sec:Algorithm1}, we present an optimization algorithm for the random access process via POMDP formulation. Then resource allocation based on the feedback and control loop is formulated in Section \ref{sec:Algorithm2}. Section \ref{sec:Simulation} discusses the simulation results. Finally, we conclude this work in Section \ref{sec:Conclusion} with future works.\

\section{Overview of Software-defined Cellular Networks with M2M Communications and Wireless Network Virtualization}\label{sec:Overview}
In this section, we first describe the architecture and RA procedure in M2M communications. Then, wireless network virtualization with M2M communications will be discussed. The functions and features of SDN will be introduced as well.\

\subsection{Random Access in M2M Communications}
\begin{figure}[!t]
\centering
\subfloat[]{\includegraphics[width=1.8in]{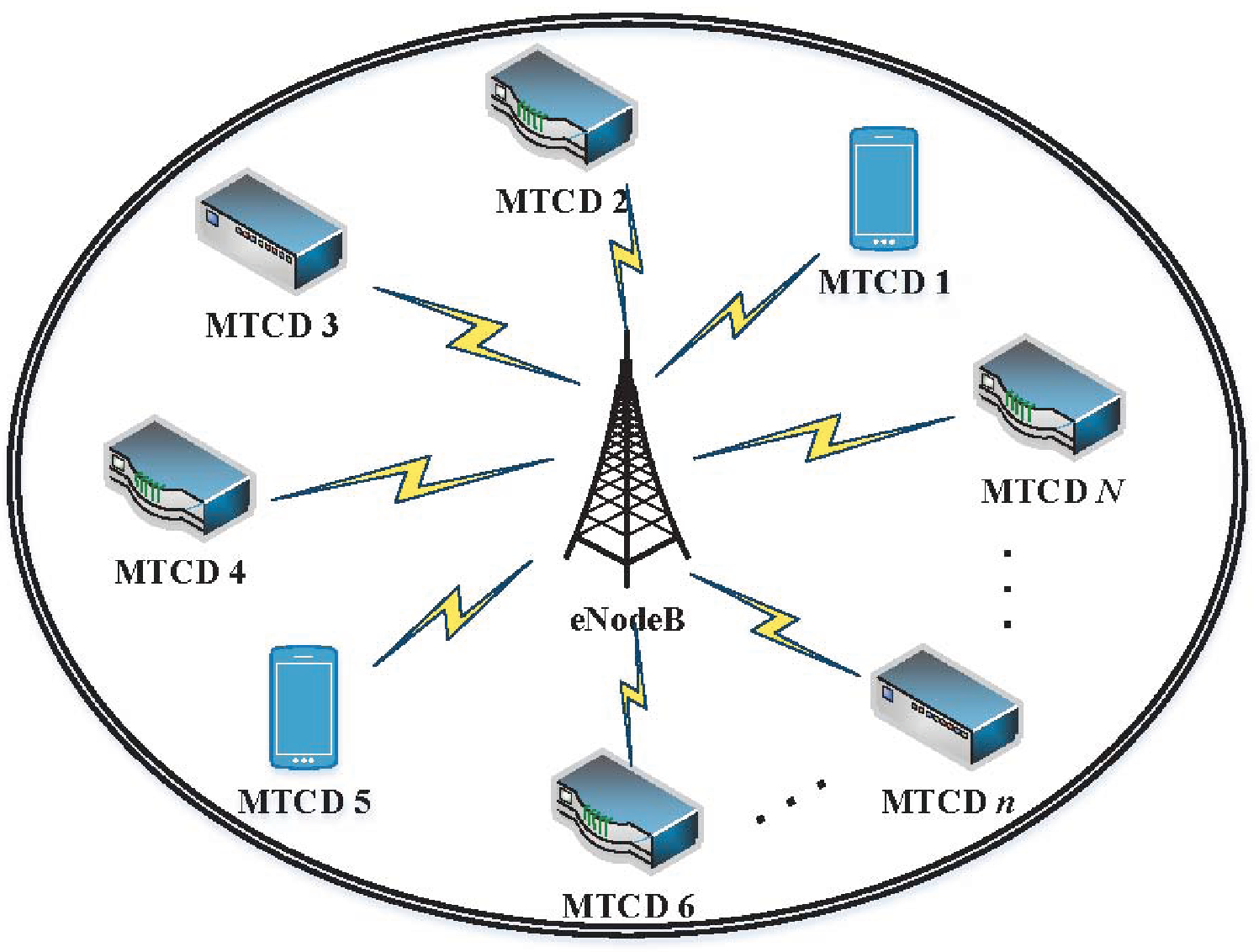}}
\subfloat[]{\includegraphics[width=1.8in]{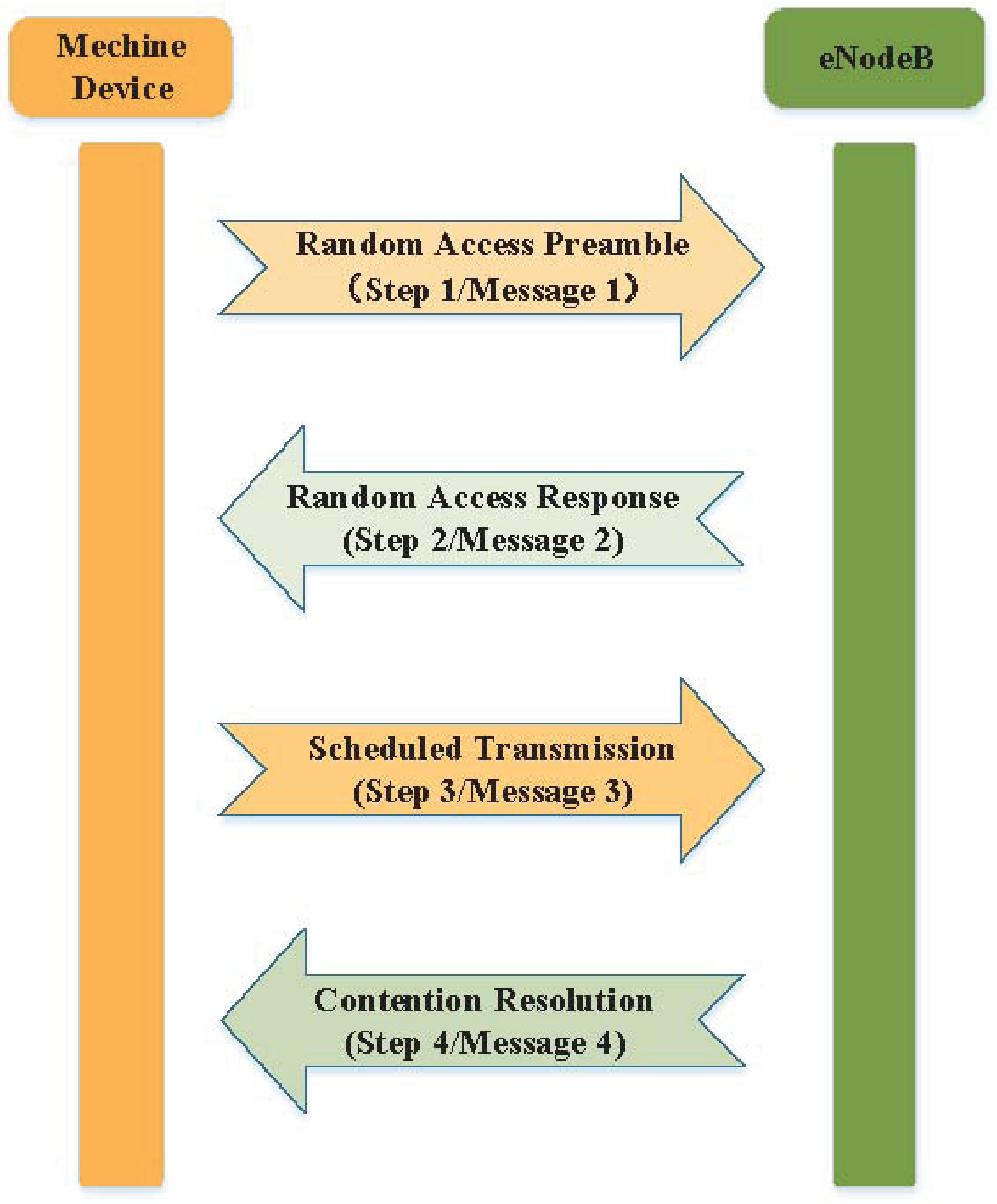}}
\caption{(a) A general architecture of an M2M communication network and (b) a random access procedure in an M2M communication network.}
\label{fig:architecture}
\end{figure}

For M2M communication, a variety of MTCDs exist in the networks, and the architecture of an M2M communication network with a single cell is described in Fig. \ref{fig:architecture}(a)~\cite{OH15}. For simplicity, mobility \cite{YL01,GYJ10} and handover \cite{MYL04} are not considered in this paper. Similar to a normal user equipment (UE), an MTCD has the ability to establish a direct link with eNodeB to access. In general, there are two different forms of random access procedure in M2M communications ~\cite{LA15}:\

\emph{Contention-based}: The MTCDs may collide when they compete for the channel access. Therefore, this may incur delay-tolerant access requests.\

\emph{Contention-free}: The MTCDs may receive the specific access resource from the eNodeB, then the MTCDs can obtain access chance with high probability of success.\

In this paper, we mainly focus on contention-based RA mechanisms. The contention-based RA procedure includes four steps before an MTCD establishes connection with an eNodeB, which is described in Fig. \ref{fig:architecture}(b)~\cite{OH15}. The four steps are described as follows~\cite{ZK15}:\

\emph{Step 1. Random Access Preamble}\

{\color{black}When an MTCD attempts to access the network, it can select a random access preamble to send to the eNodeB through RB in the RA slot. In this step, there are two cases that may occur: one is that the same preamble may be selected simultaneously by more than one MTCD; the other is that the same RB may be utilized simultaneously to send the preamble by two or more MTCDs. As a result, the eNodeB will receive the same uplink information and transmit scheduled messages, or receive messages on the same uplink resource. Then, the MTCD, which selects the same preamble or same RB with others, will experience a collision in this time slot.}\

\emph{Step 2. Random Access Response (RAR)}\

In this step, the eNodeB will decode the preamble transmitted by the MTCD. If the preamble can be decoded successfully, the eNodeB will compute an identifier, which is calculated based on the RA slot where each preamble is sent. Then, an RAR will be transmitted through the Physical Downlink Shared Channel (PDSCH) by the eNodeB. The RAR conveys the identity of the detected preamble, uplink grant for the scheduled message and the assignment of a temporary identifier.\

\emph{Step 3. Scheduled Transmission}\

The MTCD will transmit a connection request message to the eNodeB with the resources granted in Step $2$. The request message associates with the preamble transmitted in the RA slot. In this step, a Hybrid Automatic Retransmission Request (HARQ) will be transmitted. There is a special case: the preamble collision may not be detected by the eNodeB. Then more than one MTCD will use the same uplink resource to transmit message in Step 3. Consequently, a collision occurs at eNodeB. Besides, each MTCD will retransmit messages for the maximum number of retransmissions allowed before declaring access failure and scheduling a new access attempt.\

\emph{Step 4. Contention Resolution}\

If the MTCD does not receive messages from Step 2, the system will declare a failure in the contention resolution and schedule a new access attempt. If the number of retransmission reaches the maximum allowed value, the network is declared unavailable by the device and a random access problem is indicated to upper layers. Otherwise, if the scheduled message can be correctly decoded by eNodeB in this step, the eNodeB will transmit the contention resolution to the corresponding MTCD. In other words, once the MTCD receives and decodes contention resolution message successfully in Step 4, the random access procedure will be completed, then the MTCD and eNodeB can communicate with each other.\

\subsection{Wireless Network Virtualization with M2M Communications}
Wireless network virtualization has been considered as a promising technology for next generation wireless networks, and it has a very broad scope ranging from spectrum sharing, infrastructure virtualization, to air interface virtualization~\cite{LYZ15}. Based on different QoS requirements, a physical wireless network can be virtualized into several virtual wireless networks, which share the same infrastructures, radio spectrum resources and/or RBs. Therefore, wireless network virtualization will promote the development of new communication technologies (e.g., 5G and future generations) and communication schemes (e.g., M2M communications and industrial Internet).

In most existing works on M2M communications, only one class of MTCDs is considered. However, in practical networks, the QoS requirement of diverse M2M services may vary widely. To support different QoS requirements in M2M communication networks, network equipments in a physical cellular network can be virtualized into several virtual networks by the approach of wireless network virtualization. For example, according to the different functions of MTCDs and their QoS requirements, a physical cellular network can be virtualized into emergency networks, vehicular networks, industrial networks, smart grids networks \cite{BYC12} and agriculture networks, etc. These virtual networks share the same physical network to efficiently use the radio resources, computing resources, networking resource, and other resources.
\begin{figure}[!t]
\centering
\includegraphics[width=2.5in]{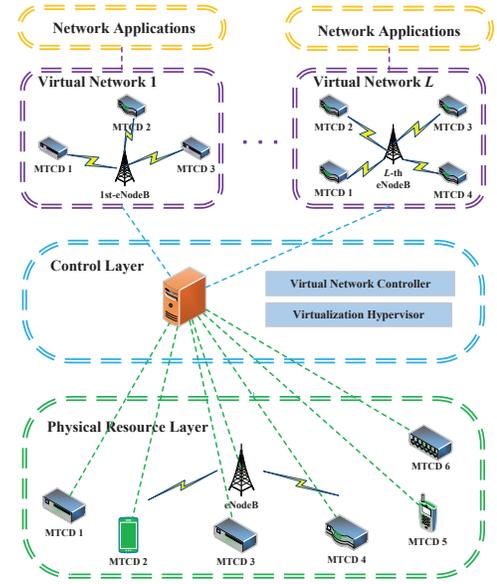}
\caption{The architecture of a software-defined cellular network with M2M communications and wireless network virtualization.}
\label{fig:model}
\end{figure}

\subsection{Software-defined Cellular Networks with M2M Communications and Wireless Network Virtualization}
To implement wireless network virtualization with M2M communications and efficient resource sharing among virtual networks, network elements in the infrastructure need to support dynamic fast (re-)configuration. However, the existing network elements and protocols were not well designed to react to changes dynamically. In addition, due to the existence of proprietary and diverse protocols and interfaces among network elements, compatibility issues arise when integrating M2M communications with wireless network virtualization. Fortunately, the SDN paradigm provides a promising platform to implement wireless network virtualization with M2M communications.

The concept of SDN was firstly proposed from the OpenFlow system by Stanford University~\cite{JK14}. The basic idea of SDN is to break vertical integration, to detach the control plane from the forwarding plane, and to introduce the ability of programming the network. In SDN, decisions are made by the ``network brain" with a global network view, which eases resource management and network optimization. Meanwhile, data plane elements become highly efficient and programmable packet forwarding devices, while the control plane elements are represented by a logically centralized single entity, the controller.  Compared with traditional networks, it is convenient to deploy and develop new applications through SDN. Moreover, with the global view of the  SDN controller, it is easier to dynamically operate, manage, and optimize the network in a timely and efficient way. SDN can greatly facilitate big data acquisition, transmission, storage, and processing \cite{CYY15}. In addition, SDN makes it easier to detect and react to security attacks \cite{YYG15}. It can also offer fine-grained virtual resource allocation based on time-varying QoS requirements and network conditions \cite{CYL15}.

An example of the software-defined cellular network with M2M communications and wireless network virtualization is depicted in  Fig. \ref{fig:model}, where the hypervisor enables the virtualization of the physical network, which is abstracted and sliced into multiple virtual networks. Through the SDN controller, dynamic resource allocation can be realized with a feedback and control loop. As can be seen in Fig. \ref{fig:model}, after virtualization, the SDN controller has a global view of each virtual network. Based on the functions and QoS requirements of different virtual networks, {\color{black}RBs can be considered as network elements \cite{XYJL12} and} dynamically adjusted and allocated by SDN controller. Besides, within a virtual network, {\color{black}RBs} that are originally used in the data transmission phase also can be dynamically adjusted by the SDN controller according to QoS requirements.

\section{System Model}\label{sec:Systemmodel}
{\color{black}In this section, we describe the system model of random access and resource allocation for software-defined cellular networks with M2M communications and wireless network virtualization in random access phase. The proposed system model can be considered as three layers. Physical resources such as RBs and eNodeBs can be virtualized as virtual resources by hypervisor, and MTCDs are mapped into the corresponding virtual network. In each virtual network, MTCDs can select proper RBs to access network with the maximum transmission rate. In addition, after each time period, if the MTCDs cannot obtain the desired transmission rate to access, especially in the virtual network with high QoS requirements, the number of RBs used in the random access phase will be dynamically allocated by the SDN controller based on different requirements of each virtual network. The detailed system model is described in the following.}\

\subsection{Layer 1. Physical Resource Layer}
\begin{figure}[!t]
\centering
\subfloat[]{\includegraphics[width=3.5in]{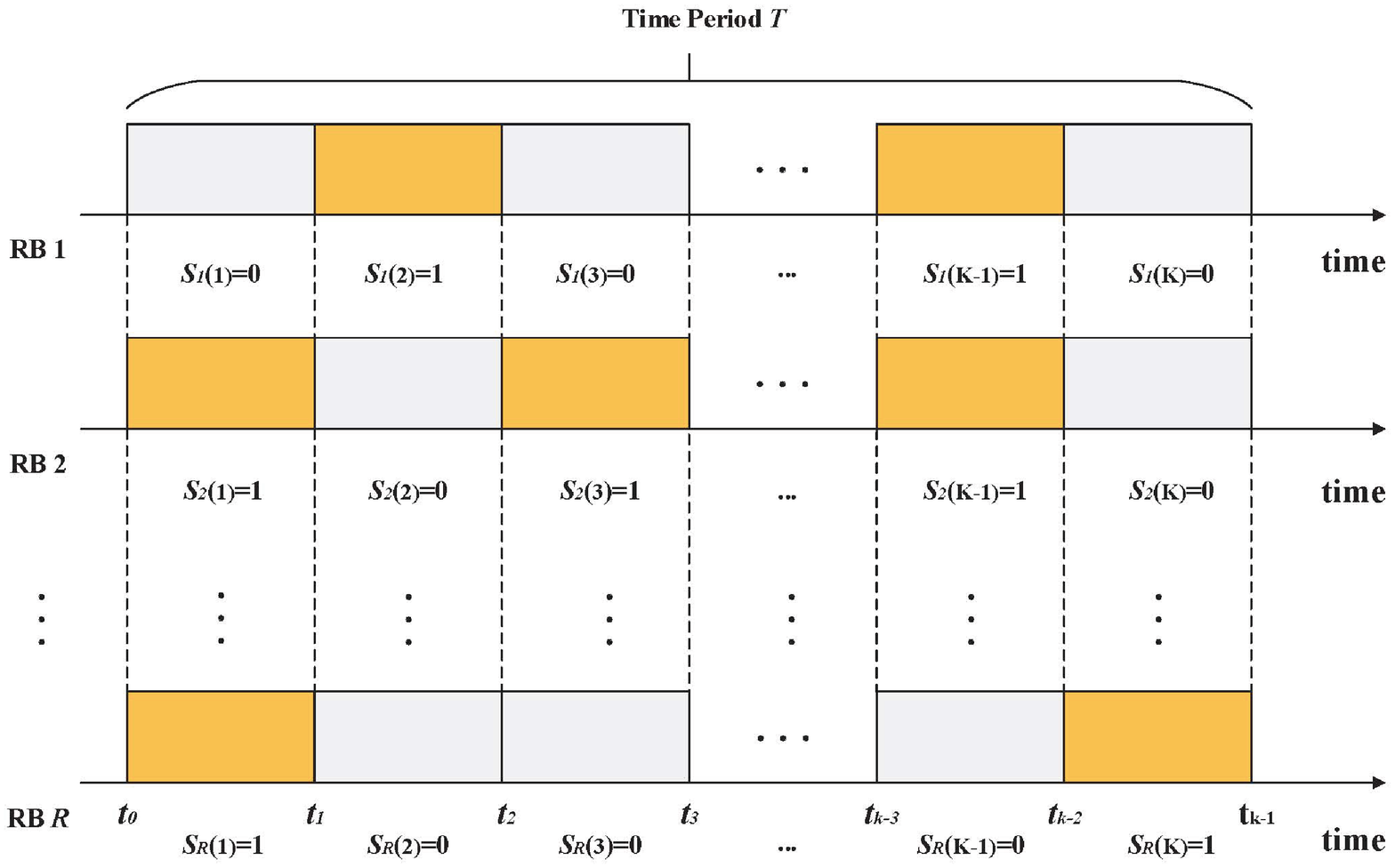}}
\hfil
\subfloat[]{\includegraphics[width=3.5in]{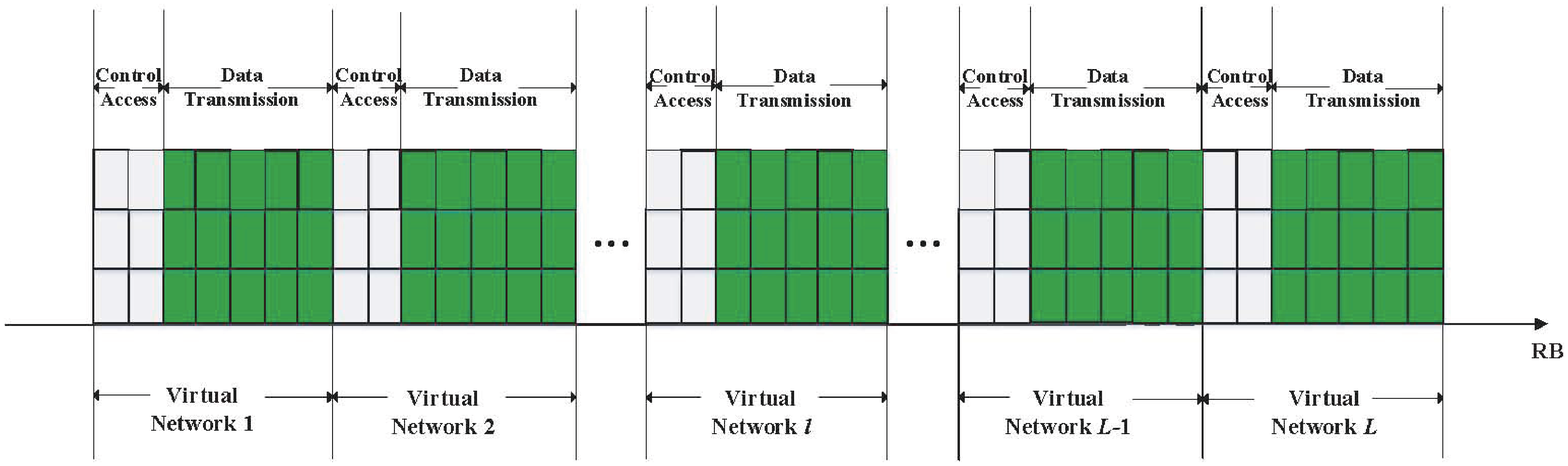}}
\caption{(a) The state of RBs in each time slot and (b) the function of RBs in different virtual networks.}
\label{fig:timeslot}
\end{figure}

\
In this layer, there are various MTCDs and eNodeBs, all of which are physical resources. As can be seen in Fig.~\ref{fig:model}, we consider the single-cell scenario with multiple MTCDs. We assume that there are $N$ MTCDs and one eNodeB in the physical cellular network. {\color{black}The time points that the MTCDs can access the eNodeB are} $t_{0}, t_{1}, ..., t_{K-1}$, where $K$ is the total number of time slots we considered, $1\leq k\leq K-1$, and each time slot is equal. It represents as $t_{k}-t_{k-1}=\delta t_k$, where $\delta t_k$ is the duration of a time slot. A time period includes the $K$ time slots, from time point $t_{0}$ to $t_{K-1}$, each time period is represented as $T_1,T_2,...,T_{y},...,T_{Y}$. Meanwhile, RBs will be offered by the eNodeB when the MTCDs attempt to access the eNodeB. We assume that the total number of RBs is $R_{total}$. The number of RBs used in the control access phase is $R$, while that used in the data transmission phase is $R^{'}$. They satisfy that $R+R^{'}=R_{total}$. Considering the RBs for the access phase, $r$ represents the $r$-th RB, where $1\leq r\leq R$. The state of each RB in one time slot can be described as idle or busy. We use the set $\bm{s_r}$ to represent the state of the $r$-th RB, and $\bm{s_{r}}=\{0,1\}$, where $0$ stands for the RB is idle while $1$ stands for the RB is busy in this time slot. The state of each RB can be described as Fig. \ref{fig:timeslot}(a).\

{\color{black} In addition, the preamble collision in the random access phase with M2M communications cannot be ignored. Let the number of available preambles be $N_p$, and the preamble selection follows a Binomial distribution with mean $\frac{1}{N_p}$~\cite{ZK15}. Therefore, the probability of preamble collisions $Pr_s$ is calculated as}

\begin{equation}
\begin{aligned}
Pr_s=binom(N,D_\chi)\cdot\left(\frac{1}{N_p}\right)\cdot{\left(1-\frac{1}{N_p}\right)}^{N-D_\chi}.
\end{aligned}
\end{equation}
{\color{black} where $N$ denotes the total number of MTCDs,} $D_\chi$ denotes the number of MTCDs that select the same preamble $\chi$, and $binom(a,b)=\frac{a!}{(b!)\times(a-b)!}$.

\subsection{Layer 2. Control Layer}\
In the proposed framework, the controller is set in this layer of the network architecture, which includes the hypervisor and SDN controller. The hypervisor is an important component in wireless network virtualization. In general, the hypervisor can be implemented at the physical eNodeB, and it provides functions to connect physical resource and virtual eNodeB~\cite{LY15}. Moreover, the hypervisor takes the responsibility of virtualizing the physical eNodeB into a number of virtual eNodeBs. Besides, the hypervisor is also responsible for scheduling the air interface resources~\cite{LYMa15}. As mentioned above, the SDN controller also plays an essential role in the proposed framework, and the network resources can be allocated dynamically by the SDN controller.\

The virtualization process can be divided into the following three steps~\cite{LY16}:

\emph{{\color{black}Step 1: Initalization}}

a) {\color{black}\emph{Slicing}: The mobile virtual network operator (MVNO) generates certain number of virtual bearers based on the functions of MTCDs and M2M communication network status.}

b) {\color{black}\emph{Define virtual resources}: The MVNO defines different virtual network functions and properties (e.g., transmission rate, delay and priority) for each virtual bearer based on the M2M network service.}

c) {\color{black}\emph{Provide virtual resources}: The MVNO delivers virtual bearers to the corresponding service providers (SPs).}

\emph{{\color{black}Step 2: Scheduling}}

a) {\color{black}\emph{Programming}: Each SP allocates appropriate number of virtual bearers to each MTCD based on its functions and QoS requirements (e.g., delay and transmission rate).}

b) {\color{black}MVNO receives the scheduling information about next potential MTCDs from SPs.}

c) {\color{black}\emph{Isolating}: The MVNO converts the properties of each virtual bearer to QoS requirements and prepares the physical resources for each MTCD.}

\emph{{\color{black}Step 3: Mapping}}

{\color{black}The MVNO allocates physical resources (e.g., eNodeB and RBs) to each MTCD based on current network status, QoS requirements or service functions {\color{black}(e.g., different requirements about access rate, delay, energy efficiency and density).}}

{\color{black}In addition, for the SDN controller in this layer, it allows a high-level abstract, to which a set of underlying network resources are automatically and dynamically mapped. Meanwhile, the eNodeBs can be implemented in a virtualized manner on general hardware coordinated and managed centrally by SDN controllers. Moreover, the SDN controller that is physically deployed on centralized servers, abstracts current resource usage and operates the network elements with intelligent strategy through standard application programming interfaces (APIs)~\cite{SG15}.} Therefore, based on the functions of SDN controller, a feedback control loop is proposed and designed in the control layer, then all of RBs offered by the eNodeB can be allocated dynamically to each virtual network by the SDN controller. According to different functions of each virtual network, the SDN controller can adjust the number of allocated RBs to optimize and improve the performance of networks. By these means, in the virtual network with M2M communications, the SDN controller will offer an efficient approach to allocate RBs for M2M communications, and will improve the performance.\

\subsection{Layer 3. Virtual Network Layer}\
As shown in Fig. \ref{fig:model}, according to different QoS requirements, the physical network will be virtualized to multiple virtual networks by hypervisor. The hypervisor takes the responsibility of mapping the physical network  with M2M communications into $L$ virtual networks. For the $l$-th $(1\leq l\leq L)$ virtual network, it includes $N_l (1\leq N_l\leq N)$ MTCDs, which have the same or similar functions. For example, one virtual network that has MTCDs for emergency services, and another virtual network that has MTCDs with utilities services. Meanwhile, in the $l$-th virtual network, the virtual eNodeB can offer all RBs to control access transmission and data transmission in the initial time slot. The numbers of RBs used for control access and data transmission are $R_l (1\leq R_l\leq R)$ and $R_{l}^{'} (1\leq R_l^{'}\leq R^{'})$, respectively, as described in Fig. \ref{fig:timeslot}(b). {\color{black}As a result, the MTCDs only need to sense and detect the RBs that belong to the corresponding virtual network, instead of scaning the whole network resources in the physical network~\cite{LY15, SJY10}.}\

{\color{black} Considering the transmission rate in the random access channel, it can intuitively reflect on the success or failure access in the random access phase~\cite{OL14}. In the proposed scheme, Shannon capacity is used for calculating the transmission rate in the random access phase~\cite{DH14}. Since the influence of preamble collisions may not avoid~\cite{OL14}, then for the $n$-th MTCD that accesses the $r$-th RB, we define $C_{l,n,r}(k)$ as the available transmission rate in the $l$-th virtual network when the $n$-th MTCD has accessed to the $r$-th RB during time slot $\delta t_k$, and it can be calculated as}\
\begin{eqnarray}
&& C_{l,n,r}(k)= \nonumber \\
&& \left\{
\begin{array}{lcl}
(1-Pr_s)B_{l,n,r} \log_2 \left\{ 1+\frac{P_{r}h_{l,n,r}}{\sigma^{2}}\right\}, \ \text{if}\ s_{r}=0,\\ \\
(1-Pr_s)B_{l,n,r} \log_2 \left\{1+\frac{P_{r}h_{l,n,r}}{\sum\limits_{n^{'}\neq n,n^{'}\in N} P_{r}h_{l,n^{'},r}+\sigma^{2}}\right\},\\   \ \ \ \ \ \ \ \ \ \ \ \ \ \ \ \ \ \ \ \ \ \ \ \ \ \ \ \ \ \ \ \ \ \ \ \ \ \ \ \ \ \ \ \ \ \text{if}\ s_{r}=1,\\
\end{array}
\right.
\end{eqnarray}
where $B_{l,n,r}$ represents the bandwidth offered by the $r$-th RB in the $l$-th virtual network, $P_{r}$ represents the transmit power  consumed by the $r$-th RB, $h_{l,n,r}$ ($h_{l,n^{'},r}$) is the channel gain when the $n$-th ($n^{'}$-th) MTCD accesses to the $r$-th RB, which includes path loss, and $\sigma^{2}$ is the system noise power.\

Due to the fact that different virtual networks include several MTCDs with different functions, we take into account that each virtual network has a different requirement of the average transmission rate when MTCDs connect to the eNodeB through RBs in the random access phase. The average transmission rate of each virtual network can be denoted as $C_{1}, C_{2}, \dots, C_{l}, \dots, C_{L}$, where $C_{1}$ represents the obtained average transmission rate of the highest level virtual network and $C_{L}$ represents the obtained average transmission rate of the lowest level virtual network.

To provide the proportional average transmission rate differentiation, the average transmission rate of the $L$ levels should be related by the expression:
\begin{eqnarray}
C_1:C_2:\ldots:C_l:\ldots:C_L\approx x_1:x_2:\ldots:x_l:\ldots:x_L,
\end{eqnarray}
where $x_l$ represents a constant weighting factor for level requirement of the $l$-th virtual network. Obviously, it satisfies that $x_{1}\geq x_{2}\geq\ldots x_{l}\geq\ldots\geq x_{L}$. For the $l$-th virtual network, the obtained average transmission rate $C_{l}$ can be calculated as
\begin{eqnarray}\label{F1}
C_l=\frac{\sum\limits_{n=1}^{N_l}\sum\limits_{r=1}^{R_l}\sum\limits_{k=1}^{K}{{C_{l,n,r}(k)}\delta t_k}}{N_l\cdot T_y}.
\end{eqnarray}

Considering different classes and QoS requirements of each virtual network, and in order to reflect relative priority and satisfy RBs allocation in different classes of virtual networks, the ratio of obtained and desired transmission rate in the $l$-th virtual network can be denoted as
\begin{eqnarray}\label{accessrate1}
\xi_l=\frac{C_l}{C_1+C_2+\ldots+C_l+\ldots+C_L},
\end{eqnarray}
\begin{eqnarray}\label{accessrate2}
\xi^{'}_l=\frac{x_l}{x_1+x_2+\ldots+x_l+\ldots+x_L},
\end{eqnarray}
where $\xi_l$ denotes the ratio of obtained transmission rate and $\xi^{'}_l$ denotes the ratio of desired transmission rate. Therefore, the gap between the ratio of desired transmission rate and obtained transmission rate can be written as $e_l=\xi^{'}_l-\xi_l$. Thus, $e_l$ is used by the SDN controller to decide the RBs adjustment and allocation in the random access phase. According to Eqs. (\ref{accessrate1}) and (\ref{accessrate2}), both $\xi_l$ and $\xi^{'}_l$ are used as the performance metrics of the feedback and control loop.\

Since both the number of RBs $R_{total}$ and the number of virtual networks $L$ are fixed, the number of RBs that could be adjusted and allocated via the control loop is limited. As a result, considering the given number of RBs and virtual networks, the ratio of transmission rate in one virtual network should be bounded. Assume the average ratio of transmission rate is $\bar{\xi}$, then the maximum ratio of transmission rate to access should satisfy
\begin{eqnarray}
\xi_{max}\leq\frac{\bar{\xi}\cdot\ln R_{total}}{\ln R_{total}-\ln L},
\end{eqnarray}
it should be noted that the ratio of the obtained and desired transmission rate must satisfy that $\xi_l\leq\xi_{max}$ and $\xi^{'}_l\leq\xi_{max}$.\


\section{Optimization of Random Access via POMDP}\label{sec:Algorithm1}
In this section, we develop a decision-theoretic approach via POMDP to optimize the random access process  discussed in Section~\ref{sec:Overview}. Then, each tuple of POMDP is described in detail, followed by the reward and optimization objective.\

\subsection{POMDP Formulation}
{\color{black}POMDP can be considered as a generalization of Markov decision process (MDP). The actions' effects on the state in a POMDP is exactly same as in an MDP. The main difference is whether or not we can observe the current state of the process. Difference from MDP, we add a set of observations to the model in a POMDP. Therefore, instead of directly observing the current state, the state gives us an observation that provides a hint about the current state~\cite{ZT07,WYS10,XYJ12}.}\

{\color{black}Considering the proposed framework with M2M communications, if the MTCDs attempt to access the network with maximum reward, they should know the RB state in each time slot. However, since the state of RBs cannot be directly and accurately obtained by MTCDs in the random access phase, they need to take action based on RB state transition and observe state. Therefore, the optimization problem of random access that can obtain the maximum reward in M2M communications is easily formulated as a POMDP formulation~\cite{LM14}.}\

\emph{1) Action Space} \

At the beginning of each slot, based on its current information state, the $n$-th MTCD will attempt to access eNodeB and determine which action to take~\cite{ZT07}. Let $\mathcal{A}$ represent the set of all available actions, and the action that can be taken by this MTCD in time slot $\delta t_k$ can be defined as
\begin{equation}
\begin{aligned}
a(k) \in \{0 (no\ access),{RB_1},{RB_2},\dots,{RB_r},\dots,{RB_{R_{l}}}\}.
\end{aligned}
\end{equation}

In set $\mathcal{A}$, $0$ represents that the MTCD will not access the eNodeB and select sleeping mode. The MTCD may select sleeping mode in many cases (e.g., all RBs are busy or the preamble that is selected by MTCD is collided before choosing RB). Let $RB_r$ represent that the MTCD will select the $r$-th RB to access to the eNodeB.\

\emph{2) State Space and Transition Probability}\

In the M2M communication network, the system state space $\mathcal{S}$ is the set of all RB states, and the state in time point $t_k$ can be denoted as $s(k)=[{s_{1}(k)}{s_{2}(k)}\dots{s_{r}(k)}\dots{s_{R_l}(k)}]$, where $s(k)\in\mathcal{S}$. Note that, the state of the $r$-th RB can be defined as
 \begin{equation}
\begin{aligned}
{s_{r}(k)} \in \{0 (idle),1(busy)\}.
\end{aligned}
\end{equation}

Assume that each RB state is discretized. The one-step transition probability of all RB states from time point $t_{k-1}$ to $t_{k}$ is denoted by
 \begin{equation}
 Pr(k)=\begin{bmatrix}
   {p_{1,1}} & {p_{1,2}} & \dots & {p_{1,j}} &  \dots {p_{1,I}}  \\
     {p_{2,1}} & {p_{2,2}} & \dots & {p_{2,j}} &  \dots {p_{2,I}}  \\
     \vdots & \vdots & &\vdots & \vdots \\
     {p_{i,1}} & {p_{i,2}} & \dots & {p_{i,j}} &  \dots {p_{i,I}}  \\
     \vdots & \vdots & &\vdots & \vdots \\
     {p_{I,1}} & {p_{I,2}} & \dots & {p_{I,j}} &  \dots {p_{I,I}} \\
\end{bmatrix}\quad.
\end{equation}\

The value of $I$ is determined by the total number of RBs and the state of each RB, leading to $I= 2^{R_{l}}$. And $p_{i,j}$ is the transition probability of all RB states from state $i$ to state $j$. Meanwhile, for the $r$-th RB, the state transition probability can be expressed as
 \begin{equation}\label{stateprobability}
\begin{aligned}
Pr_{RB_r}=Pr\{s_r(k+1)\mid s_r(k)\}.
\end{aligned}
\end{equation}

{\color{black}In fact, the state of RBs in each time slot, which includes busy or idle, may not be obtained directly and accurately. In general, in order to calculate the state transition probability of RBs with long-term statistics, the busy (or idle) RBs state can be modeled as Poisson distribution~\cite{CC10}.} In view of Poisson distribution, the transition probability of the $r$-th RB state from state $s_r(k)$ to state $s_r(k+1)$ can be calculated as
\begin{equation}
\begin{aligned}
p(s_r(k),s_r(k+1))={\frac{(\lambda)^{m}}{m!}}e^{-\lambda},
\end{aligned}
\end{equation}
where $\lambda$ is the occurred frequency of the busy (or idle) state, and $m$ is the total number of states varying from $s_r(k)$ to $s_r(k+1)$ in long-term statistics.\

\emph{3) Observation Space}\

Since it is difficult to acquire the full knowledge of each RB state, the MTCD needs to observe the  RB state based on the state transition and optimal action taken in this time slot~\cite{LM14}. Suppose that some MTCDs have decided to select RBs to access eNodeB, implying part of RBs are in busy state during this time slot. Then, the MTCD that will access the eNodeB needs to observe the RB state before making decision. Let $\theta_{r}(k)$ denote the observation state of the $r$-th RB in time slot $\delta t_k$, where $1\leq r\leq R_l$. $\theta_{r}(k)$ can be identified as
\begin{equation}
\begin{aligned}
{\theta_{r}(k)} \in \{0 (idle),1(busy)\}.
\end{aligned}
\end{equation}
Then in the time slot $\delta t_k $, the observation state can be written as ${\theta(k)}=[\theta_{1}(k)\theta_{2}(k)\dots\theta_{r}(k)\dots\theta_{R_l}(k)]$, where $\theta(k)\in\Theta$, and $\Theta$ is the set of all observation states.

{\color{black}As the $r$-th RB state transits from $s_r(k)$ to $s_r(k +1)$ under action $a(k)$, an observation state $\theta_r(k)$ is generated with the conditional probability $b^{a(k)}_{s_r(k+1),\theta_r(k)}=Pr\{\theta_r(k) \mid s_r(k+1), a(k)\}$.} Hence, the conditional probability of observation can be denoted as
\begin{eqnarray}
b^{a(k)}_{s_r(k+1),\theta_r(k)}=\left\{
\begin{array}{lll}
\epsilon, &\text {if}~a(k)=RB_r,~\theta_r(k)=0,\\
1-\epsilon, &\text {if}~a(k)=RB_r,~\theta_r(k)=1,\\
\varphi, &\text {if}~a(k)=0,~\theta_r(k)=0,\\
1-\varphi, &\text {if}~a(k)=0,~\theta_r(k)=1,\\
\end{array}
\right.
\end{eqnarray}
{\color{black} where $\epsilon$ is the the probability of false observation for mistaking the idle state or busy state when the action is to select $RB_r$, and $\varphi$ is the the probability of false observation for mistaking the idle state or busy state when the action is not to select any RB.} For the sake of simplicity, in this article, we assume that $\epsilon=\varphi$.

\emph{4) Information State}\

In POMDP formulation, information state is an important element. Although the RB state cannot be directly known by the MTCD, it can be obtained from its action decision and observation history encapsulated by the information state, i.e., a probability distribution over states, is sufficient statistics for the history, which means that the optimal decision can be made based on the information state.\

Let $\pi(k)=\{\pi_1^{k}, \pi_2^{k},\dots,\pi_{s_r(k)}^{k},\dots,\pi_{s_{R_l}(k)}^{k}\}, {s_r(k)\in\mathcal{S}}$ denote the information space, where $\pi_{s_r(k)}^{k}\in[0,1]$ is the conditional probability (given decision and observation history) that the $r$-th RB state is in $s_r(k)$ at the beginning of time slot $\delta t_k$ prior to state transition. As will be shown later, the knowledge of the system dynamics and the transition probabilities are necessary to maintain an information state.\

The information state can be easily updated after each state transition to incorporate additional step information into history. Given $\pi_{s_r(k)}^{k}$, then after taking action $a(k)$ and observing $\theta_r(k)$, the information state is updated
\begin{equation}
\pi_{s_r(k+1)}^{k+1}=\kappa\sum\limits_{s_r(k)\in\mathcal{S}}\pi_{s_r(k)}^{k}p(s_r(k),s_r(k+1))b^{a(k)}_{s_r(k+1),\theta_r(k)},
\end{equation}
where $\kappa$ is a normalizing constant and it can be calculated as
\begin{equation}
\kappa=\frac{1}{\sum\limits_{s_r(k)\in\mathcal{S}}\sum\limits_{{s_r(k+1)}\in\mathcal{S}}\pi_{s_r(k)}^{k}p(s_r(k),s_r(k+1))b^{a(k)}_{s_r(k+1),\theta_r(k)}}.
\end{equation}

Therefore, the information state is updated by using Bayes' rule at the end of each time slot as follows~\cite{LM14,SS73},
$\pi_{s_r(k+1)}^{k+1}=$
\begin{equation}
\frac{\sum\limits_{s_r(k)\in\mathcal{S}}\pi_{s_r(k)}^{k}p(s_r(k),s_r(k+1))b^{a(k)}_{s_r(k+1),\theta_r(k)}}{\sum\limits_{s_r(k)\in\mathcal{S}}\sum\limits_{{s_r(k+1)}\in\mathcal{S}}\pi_{s_r(k)}^{k}p(s_r(k),s_r(k+1))b^{a(k)}_{s_r(k+1),\theta_r(k)}}.
\end{equation}

 The information states capture all the history information at time slot $\delta t_k$. Therefore, the past actions and observations can be saved by constantly updating the information state. In other words, it is reasonable to make decisions according to the information state.\

\emph{5) Reward and Objective}\

In the $l$-th virtual network, idle RBs may be offered by eNodeB and the MTCD will choose the RB with better performance to access. By regarding the transmission rate as a reward, the maximum transmission rate offered by RB can be used for performance evaluation. Since each system state is decided by all $R_l$ RBs states, the maximum value of the transmission rate offered by RB will be taken as the reward. Hence, for each system state, the corresponding transmission rate can be denoted as
\begin{equation}
C_{l,n}(k)=\max\{C_{l,n,1}(k),\dots, C_{l,n,r}(k),\dots, C_{l,n,R_l}(k)\}.
\end{equation}

Then the optimization objective is to maximize the transmission rate that can be achieved by MTCDs in each time period. Therefore, the system reward in the proposed scheme within time slot $\delta t_k$ is originally defined as
\begin{eqnarray}
Re_{l,n}(k)=\left\{
\begin{array}{lll}
0, &\text {if there is no sensing},\\
C_{l,n}(k), &\text {otherwise},\\
\end{array}
\right.
\end{eqnarray}
and the total discounted reward $Re_{l,n}$ is
\begin{eqnarray}
Re_{l,n}=\sum\limits^{K-1}_{k=0}\beta^{K-k-1}Re_{l,n}(k),
\end{eqnarray}
where $\beta\in[0,1]$ is the discount factor, {\color{black}which represents the difference in importance between future rewards and present rewards \cite{YK07}. For instance, when $\beta$ approaches $0$, the MTCD only cares about which action will yield the largest expected total immediate rewards; when $\beta$ approaches $1$, the MTCD cares about maximizing the expected sum of future rewards.}\

{\color{black}The optimal policy $\bm{U}$ in this paper is represented as the set of behaviors $a(k)$, $0\leq k\leq K-1$, which maximizes the expected long-term total discounted reward $Re_{l,n}$ during a time period. Hence, the optimal policy is represented as follows,
\begin{eqnarray}
\bm{U}=\{a(k)\}=\arg\max\limits_{a(k)\in \mathcal{A}}\frac{1}{K}\left[\sum\limits^{K-1}_{k=0}\beta^{K-k-1}Re_{l,n}(k)\right].
\end{eqnarray}
Since $Re_{l,n}(k)$ depends on $a(k)$, the optimal action $a(k)$ that makes the MTCD to obtain the maximum rewards is unique in each time slot.

\subsection{Solving the POMDP Problem}
In this subsection, we derive an optimal policy for selecting RB by the MTCD to maximize the transmission rate in M2M communication network based on POMDP formulation, depends on the state of each RB. In order to solve the formulation, a dynamic programming method is used in this paper.\

Let $J_{k}(\pi(k))$ be the maximum expected reward that can be obtained from time slot $\delta t_{k} (1\leq k\leq K)$, given the information state $\pi(k)$ at the beginning of time slot $\delta t_{k}$. Assuming that the MTCD that attempts to access the $r$-th RB makes action $a(k)$ and observes state $\theta_r(k)$, the reward can be accumulated starting from time slot $\delta t_{k}$. It should be noticed that the reward includes two parts~\cite{LY10}: one is the immediate reward $Re_{l,n}$, the other is the maximum expected future reward $J_{k+1}(\pi(k+1))$ starting from time slot $\delta t_{k+1}$, given the information state $\pi(k+1)$. As a result, the optimal policy of random access can be calculated as
\begin{equation}\label{totalreward}
\begin{split}
J_{k}(\pi(k))=\max\limits_{a(k)\in\mathcal{A}}\sum\limits_{s_r(k)\in \mathcal{S}}\sum\limits_{s_r(k+1)\in \mathcal{S}}\pi_{{s}_r(k)}^{k}p(s_r(k),s_r(k+1))\\ \sum\limits_{s_r(k+1)\in \mathcal{S}}b^{a(k)}_{s_r(k+1),\theta_r(k)}[Re_{l,n}(k)+J_{k+1}(\pi(k+1))],\\ \forall 1\leq k\leq K-1.
\end{split}
\end{equation}
From Eq. (\ref{totalreward}), it can be noticed that the policy of random access with M2M communications in system network will affect the total reward in two ways: firstly, how to obtain the immediate reward; secondly, how to transform and select the information state that determines the future reward.\

{\color{black}Moreover, Sondik and Cassandra showed that the value function with finite horizon is Piecewise Linear and Convex (PWLC)~\cite{SS73}. The value function of infinite horizon POMDP is not always PWLC, but can be approximated with the value function of a large enough finite horizon POMDP~\cite{SY16}. The piecewise theory is useful since the value function can be represented by a finite set of vectors, this means that the value function can be represented with a set of linear segments~\cite{SY16,ZY12}. Therefore, the domain of $J_{k}(\pi(k))$ can be partitioned into a finite number of convex regions $F_1(k), F_2(k),\ldots, F_M(k)$~\cite{LY10}, and it can be written simply as
\begin{equation}
\begin{split}
J_{k}(\pi(k))=\max\limits_{m}\sum\limits_{s_r(k)\in \mathcal{S}}\pi_{s_r(k)}^{k}\alpha_{s_r(k)}^m(k),
\end{split}
\end{equation}
where $\alpha_{s_r(k)}^m(k)\in\{\alpha_{s_r(k)}^0(k), \alpha_{s_r(k)}^1(k), \ldots\, \alpha_{s_r(k)}^M(k)\}$ is associated with each region $F_m(k)$. The set of $\alpha$-vectors represents the coefficients of one of the linear pieces of a piecewise linear function~\cite{SY16,ZY12}. These piecewise linear functions can represent the value functions for each step in the finite horizon POMDP problem. We only need to find the vector that has the highest dot product with the information state to determine what action to take. Due to limited space, the detailed explanations and corresponding programming codes of these algorithms can be found in~\cite{Tony}. The code of the incremental pruning algorithm from~\cite{Tony} will be modified and used in our examples.}


\section{Resource Allocation via Feedback and Control}\label{sec:Algorithm2}
In this section, we will present a strategy of feedback and control to allocate RBs that are used in the random access phase in a virtual network by the SDN controller. After that, we will give a detailed design method of the control loop, and propose a novel approach for RBs allocation and adjustment with M2M communications.\

\subsection{Resource Allocation with SDN Controller}
In each virtual network, after a random access process through POMDP in one time slot, the MTCD will make a decision to access eNodeB via the $r$-th RB or not. However, if one MTCD decides not to access eNodeB in excessive time slots, especially in the virtual network with high QoS requirement (e.g., in an emergency network), it will affect the system performance and QoS requirement. In addition, if the obtained transmission rate cannot reach the desired one in the access phase, it will also degrade the system performance. In other words, if there are no enough available RBs from the virtual eNodeB, the MTCDs cannot access in time, resulting in QoS degradation and transmission failure. In traditional M2M communications, if they cannot access eNodeB or access it with low transmission rate for a long time, it will degrade performance of M2M communications.

In the proposed scheme, the virtual network with the SDN controller can solve this problem through the control loop. According to the class of virtual networks, the RBs will be allocated dynamically between the control access phase and data transmission phase or between virtual networks. {\color{black}In particular, if the obtained transmission rate cannot satisfy the desired one after a time period, the SDN controller will allocate more RBs (originally used in other virtual networks or data transmission phase) to the random access phase, ensuring a better transmission rate to access eNodeB for each MTCD.} On the contrary, if the obtained average transmission rate can reach the desired one, the SDN controller will not adjust and allocate RBs. Since the state of RB in each time slot is not fixed, the obtained transmission rate is also not constant. According to the gap of ratio between the obtained and desired transmission rate after each time period, the process of resource allocation through the SDN controller is flexible and dynamic.\

\begin{figure}[!t]
\centering
\subfloat[]{\includegraphics[width=3.5in]{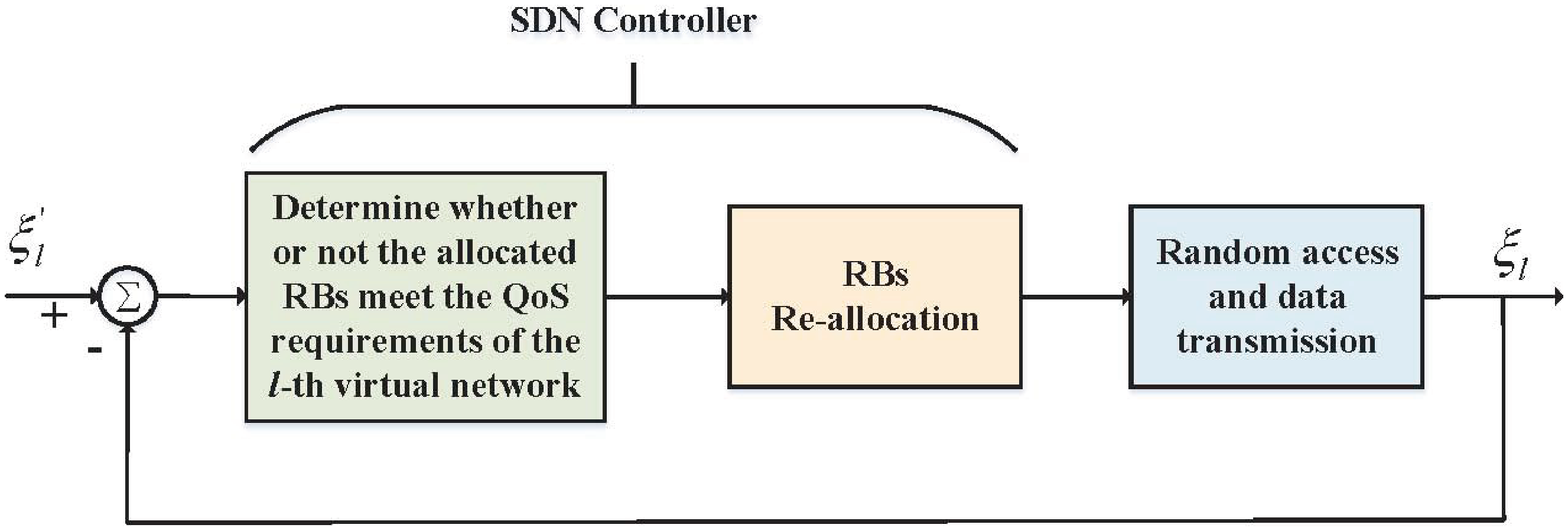}}
\hfil
\subfloat[]{\includegraphics[width=3.5in]{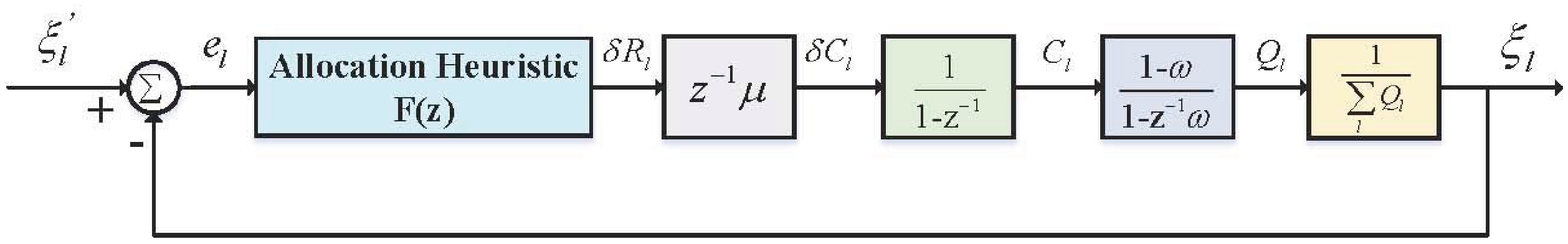}}
\caption{(a) The feedback and control loop for RBs allocation. (b) $z$-transform used in the feedback and control loop with M2M communications.}
\label{fig:allocation}
\end{figure}

After a time period, if the obtained average transmission rate cannot reach the desired one (or minimum average transmission rate), a virtual network needs a feedback mechanism to adjust RBs allocation in the access phase based on the gap of ratio between the obtained and desired average transmission rate. As mentioned above, due to the constant number of RBs in the access phase, a feedback and  control strategy to dynamically allocate RBs is necessary. With this strategy, the RBs originally used in the data transmission phase or distributed other virtual networks will be transferred to the random access phase~\cite{MS14}. The proposed feedback mechanism is depicted in Fig. \ref{fig:allocation}(a).\

In each virtual network, the number of RBs that are assigned by the virtual eNodeB is fixed in the access phase. For the $l$-th virtual network, it is expressed as $R_l$, and $\sum\limits_{l=1}^{L} R_l=R$. Based on the ratio of obtained and desired transmission rate, which is calculated by $\xi_l$ and $\xi_l^{'}$, the RB allocation algorithm through the control loop can be developed. With the proposed algorithm, the objective converts to adjust the RB allocation between random access phase and data transmission phase, or among virtual networks by the SDN controller. Consequently, a desired transmission rate in the access phase can be reached. Moreover, let the gap of ratio between the obtained and desired transmission rate after a time period $T_{y}$ be $e_l[T_{y}]$. In order to compute the reassignment number of RBs $\delta R_l[T_{y}]$, the SDN controller will utilize a linear function $f(e_l)$ and compute $\delta R_l[T_{y}]$ as follows
\begin{eqnarray}\label{numbergap1}
\forall l: \delta R_l[T_{y}]=f(e_l[T_{y}]),
\end{eqnarray}
and the number of RBs in time period $T_{y}$ is adjusted as
\begin{eqnarray}\label{numbergap2}
\forall l: R_l[T_{y}]=R_l[T_{y-1}]+\delta R_l[T_{y}].
\end{eqnarray}

According to Eqs. (\ref{numbergap1}) and (\ref{numbergap2}), the allocation strategy is concluded as: if the correction $\delta R_l[T_{y}]$ is positive, the number of RBs allocated to $l$-th virtual network in the access phase is increased by $\mid\delta R_l[T_{y}]\mid$; otherwise, it will be decreased by that number. The detailed design of the feedback and control loop will be introduced and function $f(e_l)$ will be given.

\subsection{Feedback and Control Loop Design}
In this subsection, a control loop-based model is proposed in order to design function $f(e_l)$. In essence, an approximate linear model is alternative to simplify the design of the feedback control mechanism, due to the nonlinear relationship between the adjustment number of RBs and the gap of ratio between the obtained and desired transmission rate. Nevertheless, according to the control theory, linearization is a well-known technique to solve the nonlinear problems~\cite{LA04}. Due to the linear allocation behavior, the relationship between the variation of average transmission rate and the adjustment number of RBs is approximatively proportional and can be described as
\begin{eqnarray}
\delta C_l[T_{y}]\approx\mu\delta R_l[T_{y-1}],
\end{eqnarray}
where $\mu$ is a proportionality coefficient. Then the obtained transmission rate and the variation of transmission rate should satisfy
\begin{eqnarray}
C_l[T_y]=C_l[T_{y-1}]+\delta C_l[T_y].
\end{eqnarray}

Considering Eq. (\ref{accessrate1}), it is worth noting that the obtained average transmission rate $C_l[T_y]$ might have a large standard deviation, compared with the expected value except that the time period is sufficiency large. It means that directly using $C_l[T_y]$ in the  feedback loop by the SDN controller will lead to a significant negative influence. In order to solve this problem, a low pass filter will be applied in the feedback loop. By letting $Q_l[T_y]$ be the output of $C_l[T_y]$ through the smooth filter, it follows that
\begin{eqnarray}
Q_l[T_y]=\omega Q_l[T_{y-1}]+(1-\omega)C_l[T_y],
\end{eqnarray}
where $\omega$ is a factor and satisfies that $0<\omega<1$.\

The $z$-transform is classic technique widely used in the control literature~\cite{LA04}. Through transformation and equivalent algebraic equations, the process can be easily manipulated. As can be seen in Fig. \ref{fig:allocation}(b), the control loop shows the process and relationship in the $z$-transform. The function $f$ with respect to the RB number adjustment through $z$-transform can be expressed as $F(z)$.\

According to Fig. \ref{fig:allocation}, $\xi_l$ can be denoted as
\begin{eqnarray}
\xi_l[T_y]=e_lF(z)G(z),
\end{eqnarray}
where
\begin{eqnarray}\label{ztransform1}
G(z)=\frac{z^{-1}\mu(1-\omega)}{(1-z^{-1})(1-z^{-1}\omega)\sum\limits_{l=1}^{L}Q_l}.
\end{eqnarray}
Then, by substituting for $e_l$ and using simple algebraic manipulation, the relationship between the ratio of obtained and desired transmission rate can be represented as
\begin{eqnarray}\label{ztransform2}
\xi_l=\frac{F(z)G(z)}{1+F(z)G(z)}\xi^{'}_l.
\end{eqnarray}

In order to design the RBs number allocation consistent with desired behavior of the closed loop, $\xi_l[T_y]$ should follow $\xi^{'}_l[T_y]$ within one time period. In other words, $\xi_l[T_y]=\xi^{'}_l[T_{y-1}]$ must hold. In the $z$-transform, the corresponding condition can be represented as
\begin{eqnarray}
\xi_l=z^{-1}\xi^{'}_l.
\end{eqnarray}

As a result, according to Eqs. (\ref{ztransform1}) and (\ref{ztransform2}), the design equation is
\begin{eqnarray}
\frac{F(z)G(z)}{1+F(z)G(z)}=z^{-1}.
\end{eqnarray}
After some manipulations, it results in
\begin{eqnarray}\label{ztransform3}
F(z)=\frac{z^{-1}}{(1-z^{-1})G(z)}.
\end{eqnarray}
Meanwhile, substituting for $G(z)$ into Eq. (\ref{ztransform3}), $F(z)$ is represented as
\begin{eqnarray}
F(z)=\frac{(1-z^{-1}\omega)\sum\limits_{l=1}^{L}Q_l}{\mu(1-\omega)}.
\end{eqnarray}

At last, according to the $z$-inverse transform, the  number of RBs re-assigned from the data transmission phase or other virtual networks can be calculated as
\begin{eqnarray}\label{ztransform4}
\delta R_l[T_y]=f(e_l)=\frac{\sum\limits_{l=1}^{L}Q_l}{\mu(1-\omega)}(e_l[T_y]-\omega e_l[T_{y-1}]).
\end{eqnarray}

In the $l$-th virtual network, the number of RBs that needs to be allocated and adjusted after each time period is given by Eq. (\ref{ztransform4}). The SDN controller will adjust the number of RBs based on $\delta R_l[T_y]$ from the data transmission phase or other virtual networks to the access phase so that the transmission rate in the $l$-th virtual network (especially that with high QoS requirements) can be increased and ensured.\

\section{Simulation Results and Discussions}\label{sec:Simulation}
In this section, simulation results are presented to evaluate the performance of both the proposed random access optimization algorithm modeled by POMDP and the RB allocation algorithm realized by the control loop. The considered network scenario is depicted in Fig. \ref{fig:model}. {\color{black}The physical network is considered as a single-cell scenario with a radius of 1 KM, which includes one eNodeB and $N=50$ randomly distributed MTCDs.} The MTCDs are assumed as static and deployed in the outdoor environment. According to~\cite{HR12}, the outdoor path loss model with M2M communications can be given as $8+37.6\log_{10}(d(m))$. Meanwhile, the eNodeB offers $R=25$ RBs for MTCDs and the number of available preambles is $64$ in the random access phase. The physical network can be sliced into $L$ virtual networks according to the functions and requirements of MTCDs by a hypervisor, and we assume $L=5$ in the simulations. For each virtual network, it consists of one virtual eNodeB and several MTCDs. The number of MTCDs in each virtual network, however, will be varied in different simulation scenarios. The choice of the total number of time slot in the dynamic programming depends on the convergence rate of the POMDP program, which is affected by the state-transition probabilities, observation probabilities, and value functions~\cite{LY10,L91}. In the initial time slot, RBs will be allocated equally to each virtual eNodeB, and each virtual eNodeB will be allocated $5$ RBs. Each time period includes $100$ time slots, and the number of time periods $T_y$ is assumed to be $5$ or $10$. {\color{black}For the SDN controller, we use the Opendaylight SDN controller in the simulation.} Other simulation parameters will be given in different subsections.\

We study the impacts of following parameters: 1) number of RBs, 2) probability of false observation, 3) different time periods, 4) number of MTCDs in each virtual network and 5) different classes of virtual networks. We use the following  metrics to measure the performance of the proposed algorithm: (i) received average reward (transmission rate), (ii) gap of ratio between the obtained and desired transmission rate and (iii) adjustment number of RBs in each time period. The performance evaluation and comparison will be given in different aspects.\

\subsection{Performance Improvement by POMDP and Control Loop Optimization}
In this subsection, to verify the performance improvement via the proposed scheme with POMDP and control loop, we focus on the transmission rate with different simulation parameters. The virtual network with the highest QoS requirements will be selected. The state transition probability of RB can be acquired by long-term statistics. For an RB in the virtual network, the state transition matrix is constructed by probability. The probability that the RB remains in the idle state is $Pr\{s_r(k+1)=0\mid s_r(k)=0\}=0.9$, the probability that RB remains in the busy state $Pr\{s_r(k+1)=1\mid s_r(k)=1\}=0.05$, the probability that RB transits from busy to idle state is $Pr\{s_r(k+1)=0\mid s_r(k)=1\}=0.95$, and the probability that RB transits from the idle to busy state is $Pr\{s_r(k+1)=1\mid s_r(k)=0\}=0.1$. The probability of false observation ranges from $\epsilon=\varphi=0.1$ to $\epsilon=\varphi=0.8$ in different simulation environments. The discount factor $\beta=0.8$. Additionally, the available transmission bandwidth and transmit power of virtual eNodeB is $10$ MHz and $20$ dBm, respectively~\cite{CY16}. The above parameters are widely used in the existing literature~\cite{LM14}. Moreover, we suppose that the ratio of transmission rate in the virtual network with the highest QoS requirements must achieve $ 75\%$ among all virtual networks, {\color{black}and the probability of preamble collisions is set as $Pr_s=0$.}\

We  consider two different traffic scenarios: homogeneous and heterogeneous traffic scenarios. For the homogeneous traffic scenario, MTCDs will be distributed uniformly, with $N_l=10$ ($l=1,2, \dots, 5$). For the heterogeneous traffic scenario, $N_1=30$ and $N_l=5$ ($l=2, 3, \dots, 5$). Meanwhile, for performance comparison, three other schemes are also evaluated, i.e., the proposed scheme via POMDP without (w.o.) control loop, the scheme without observation and control loop, and the scheme with perfect knowledge (the RB state perfectly known)~\cite{LY10}.\

Fig. \ref{fig:RewardRB} shows the reward with different numbers of RBs in two different simulation environments. The reward of each scheme increases with the growth of the number of RBs. As can be seen in Fig. \ref{fig:RewardRB}(a), with only one RB, there is little difference between the proposed schemes via POMDP without the control loop and the existing scheme, since there is no decision flexibility. However, for the proposed scheme via POMDP and the control loop, the transmission performance is improved significantly. The reason is that the SDN controller can adjust the number of RBs to meet the QoS requirement. With the increasing number of RBs, especially, when the number of RBs reaches $5$, the proposed scheme via POMDP is more prominent than other schemes, since more RBs can be offered and more selections can be made.\

\begin{figure*}[!t]
\centering
\subfloat[]{\includegraphics[width=3in]{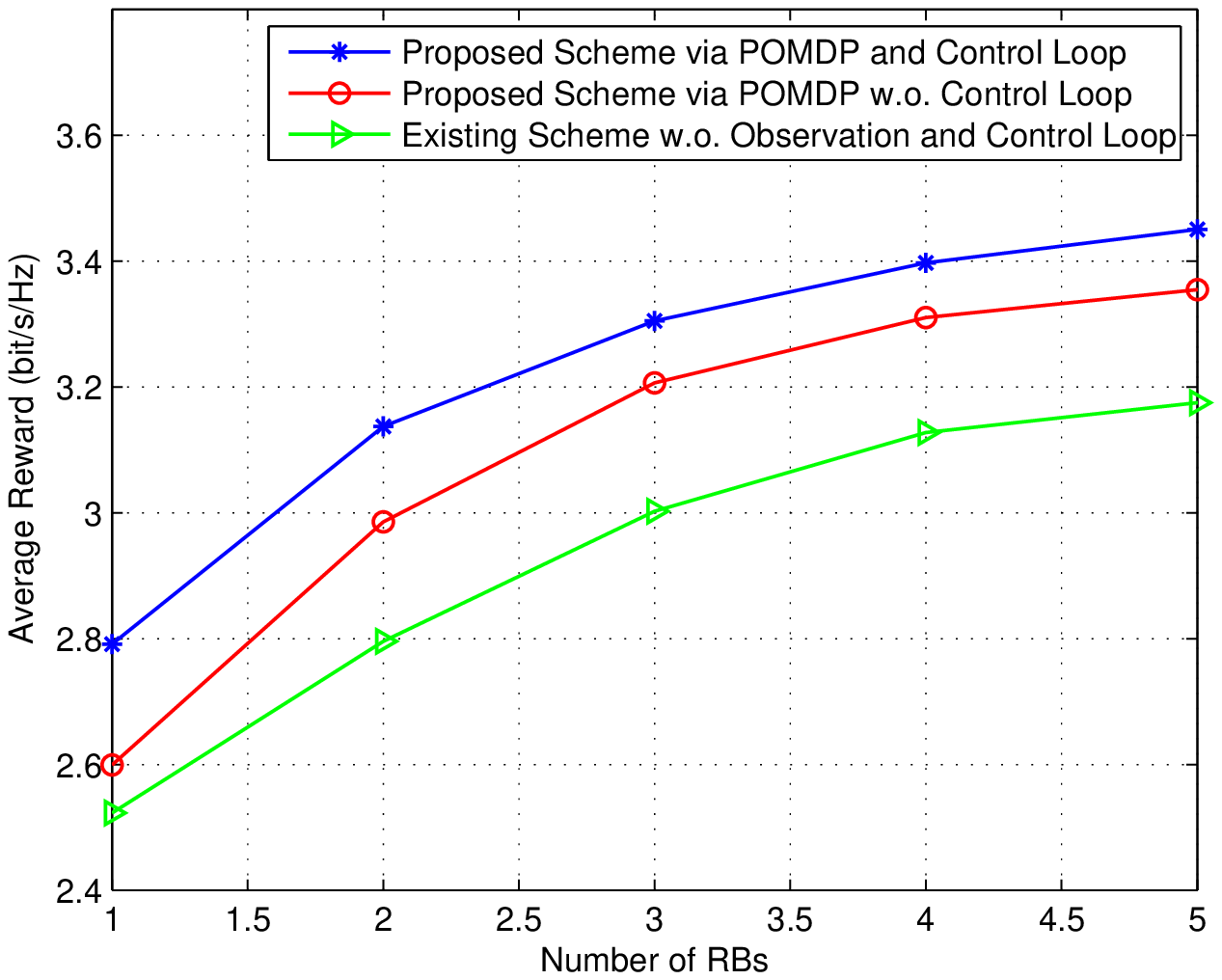}}
\subfloat[]{\includegraphics[width=3in]{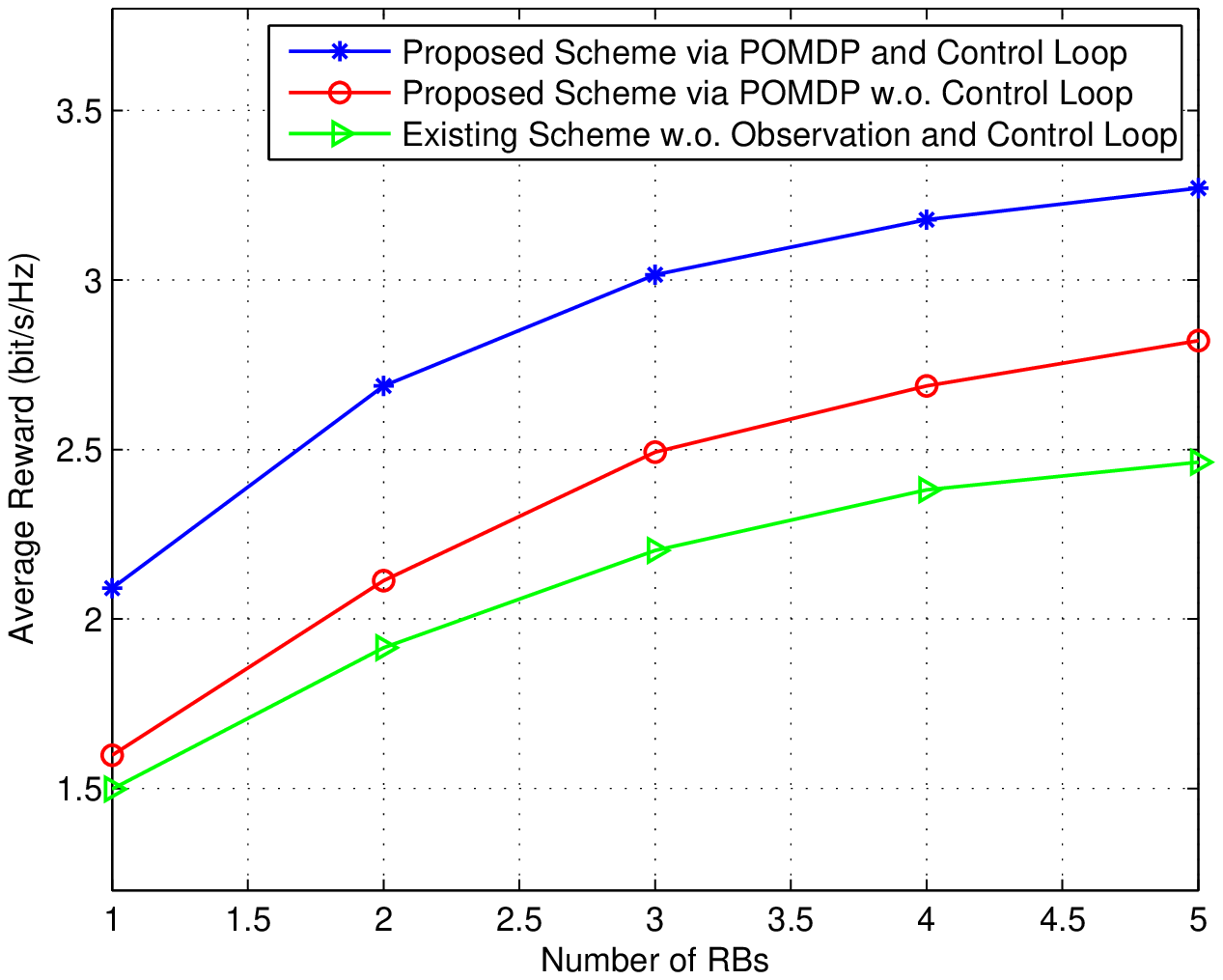}}
\caption{Average reward with different numbers of RBs in the (a) homogeneous traffic scenario and (b)  heterogeneous traffic scenario.}
\label{fig:RewardRB}
\end{figure*}

In the heterogeneous traffic scenario, MTCDs will not be distributed uniformly. Instead, more MTCDs will be distributed in the highest class virtual network. Compared with the homogeneous traffic scenario, the same trend can be found in Fig. \ref{fig:RewardRB}(b) with a more obvious advantage under the proposed scheme with the control loop. Due to the boost of MTCDs in the highest class virtual network, RBs will compete more intensely in the access phase. {\color{black}Then, the SDN controller will adjust the number of RBs to alleviate resource shortage}. The advantage of the control loop in the heterogeneous traffic scenario is also demonstrated in Fig. \ref{fig:RewardRB}(b), and the average reward in the proposed scheme via POMDP and the control loop is much higher than other schemes without the control loop. The performance improvement with the control loop in the heterogeneous traffic scenario is more significant than that in the homogeneous traffic scenario.\

Fig. \ref{fig:Rewardobserve} depicts the variation of the average reward with different probabilities of false observation. To compare the performance, we select the proposed scheme via POMDP without the control loop and the existing scheme with perfect knowledge through various observation probability to verify the performance of the proposed scheme. We also consider both homogeneous and heterogeneous traffic scenarios for ensuring fairness. From Fig. \ref{fig:Rewardobserve}, it can be seen that perfect decisions can be made when perfect knowledge of the RB state is available. However, the system cannot reach an error-free observation. Therefore, the existing scheme with perfect knowledge is only ideal and taken as an upper bound of the proposed scheme. In both Fig. \ref{fig:Rewardobserve}(a) and Fig. \ref{fig:Rewardobserve}(b), it can be easily seen that the average reward in the proposed scheme degrades with the increasing probability of false observation. When the probability of false observation keeps in low values, for instance, $\epsilon=\varphi=0.1$, the proposed methodology with POMDP and the control loop will be close to the existing scheme with perfect knowledge. However, with a higher value, such as $\epsilon=\varphi=0.8$, the performance in the proposed scheme degrades obviously. The reason is that MTCDs have to give up or select RBs with poor performance to access when the probability of false observation reaches high value resulting in lower average reward.\

\begin{figure*}[!t]
\centering
\subfloat[]{\includegraphics[width=3in]{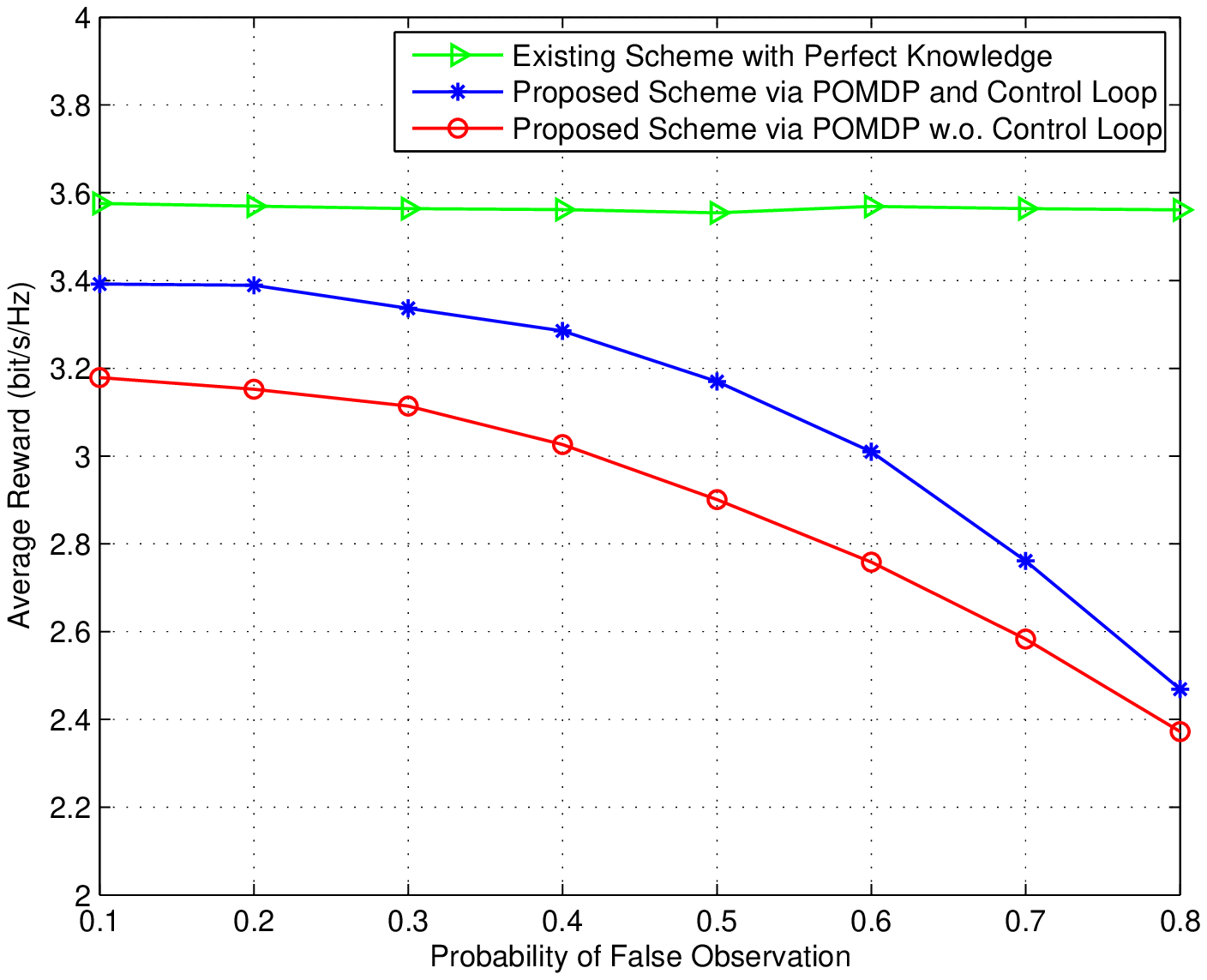}}
\subfloat[]{\includegraphics[width=3in]{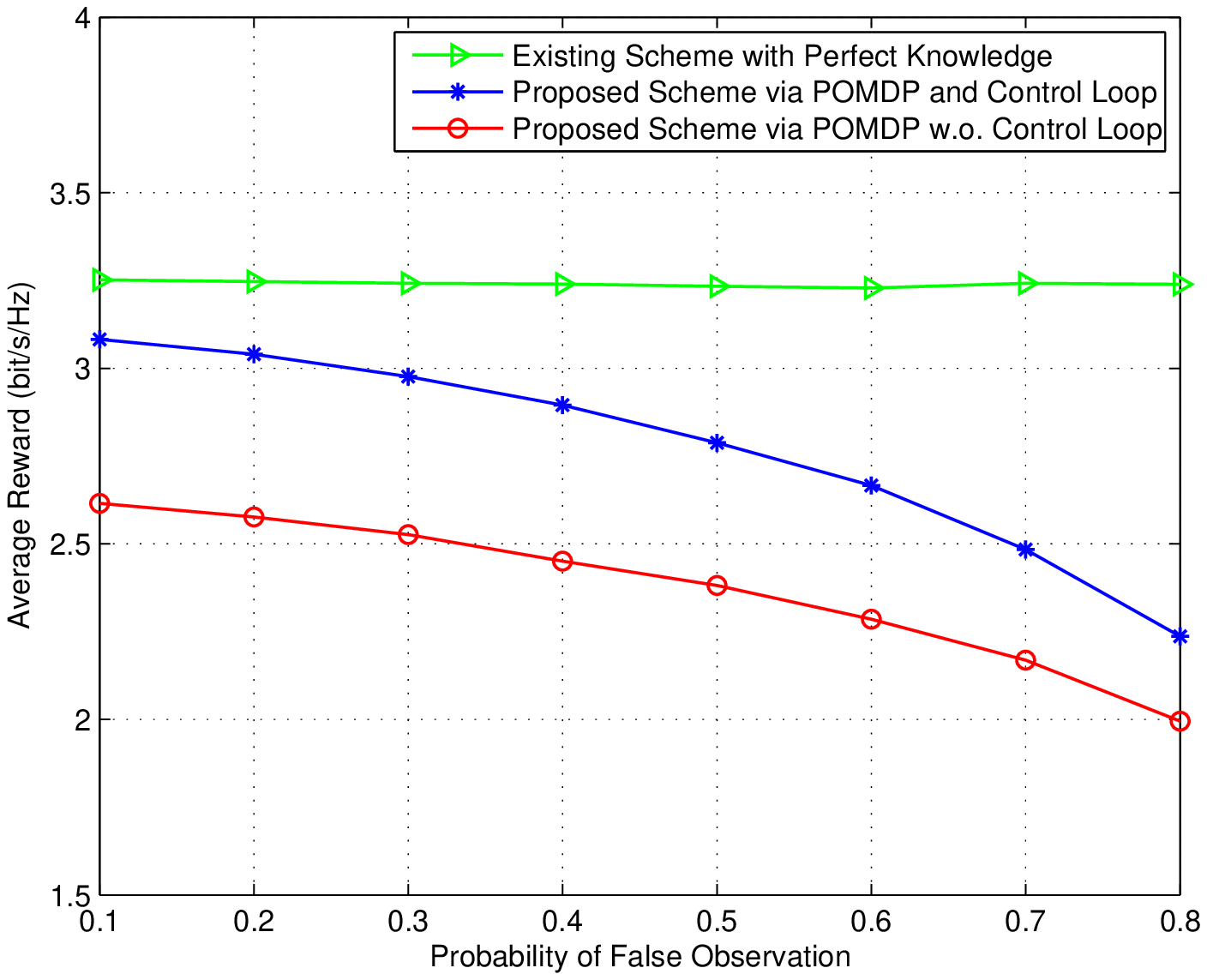}}
\caption{Average reward with different probabilities of false observation in the  (a) homogeneous traffic scenario and (b) the heterogeneous traffic scenario.}
\label{fig:Rewardobserve}
\end{figure*}

Comparing the homogeneous traffic scenario with the heterogeneous traffic scenario, we can observe that, although the variation tendency of the average reward is consistent, there are several differences between two traffic scenarios. Fig. \ref{fig:Rewardobserve}(a) and Fig. \ref{fig:Rewardobserve}(b) depict the system reward in the homogeneous and heterogeneous traffic scenarios, respectively. The performance improvement with the control loop in the heterogeneous traffic scenario is larger than that in the homogeneous traffic scenario. When the probability of false observation $\epsilon=\varphi=0.1$, the reward increases approximately $0.2$ bit/s/Hz by using the control loop in the homogeneous traffic scenario. However, under the same condition, the reward increases about $0.6$ bit/s/Hz by using the  control loop in the  heterogeneous traffic scenario, which is nearly two times higher than that in the homogeneous traffic scenario. Besides, when $\epsilon$ and $\varphi$ range from $0.1$ to $0.8$, the gap of average reward between the proposed scheme with the control loop and the proposed scheme without the control loop is still higher in the heterogeneous traffic scenario. The reason is that the number of RBs can be efficiently adjusted by the SDN controller, especially in virtual network with more RBs.\

 \subsection{RBs Adjustment and Allocation via Feedback and Control}
In this subsection, we will study the effect of RB allocation and adjustment when the control loop is added into the network. To ensure the comparison fairness, we select two virtual networks with different classes: the first virtual network with the highest class and the fifth virtual network with the lowest class. The available transmission bandwidth offered by virtual eNodeB in the first and the fifth virtual network is $10$ MHz and $5$ MHz, respectively. The transmit power is $20$ dBm in both virtual networks. Channel gains also consider the path loss. Moreover, the weighting factor is set as $x_1:x_5=3:1$. In addition, factor $\omega$ is $0.8$, the proportionality coefficient $\mu$ is $2$, {\color{black}and the probability of preamble collisions is set as $Pr_s=0$.}\
\begin{figure*}[!t]
\centering
\subfloat[]{\includegraphics[width=3in]{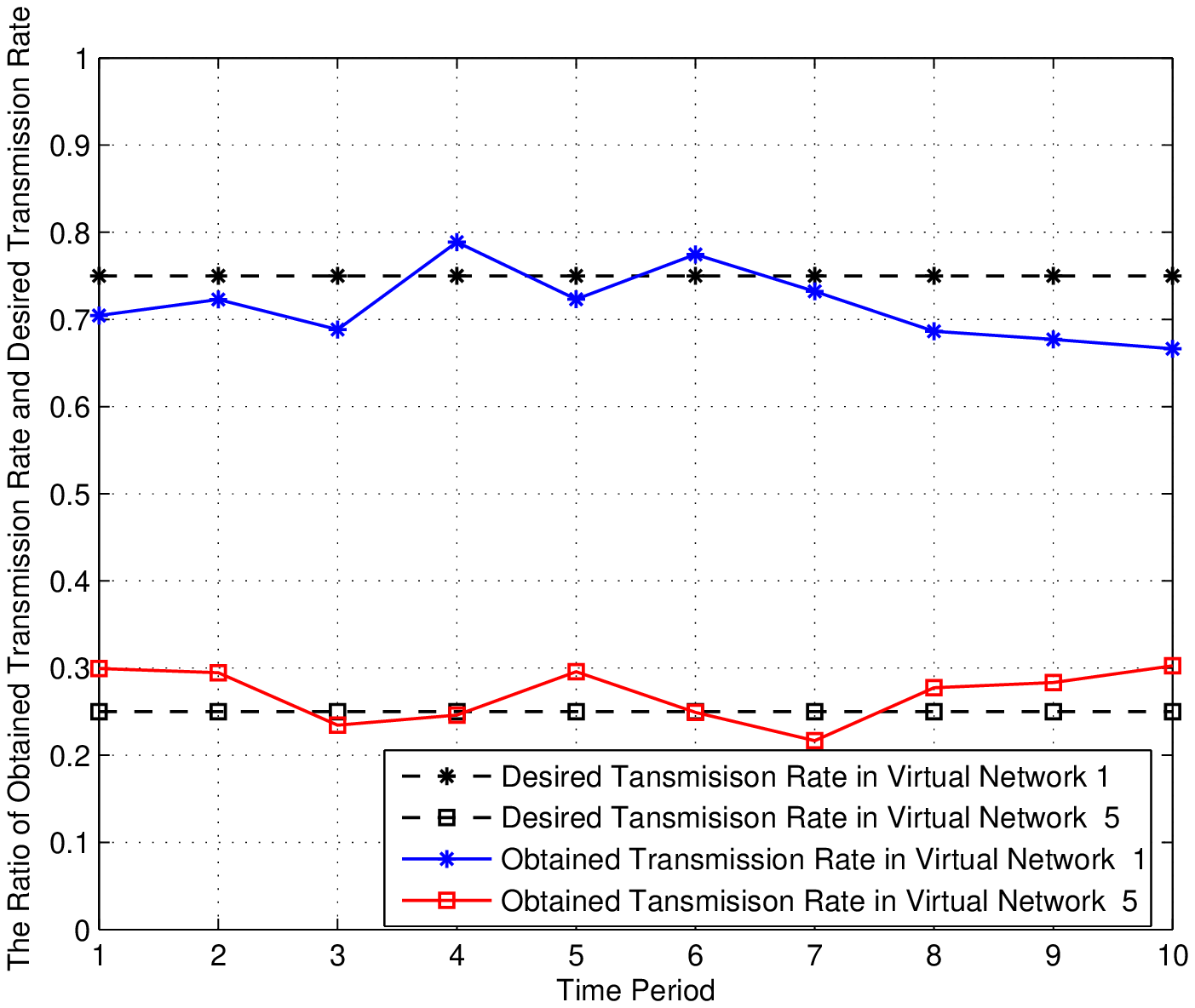}}
\subfloat[]{\includegraphics[width=3in]{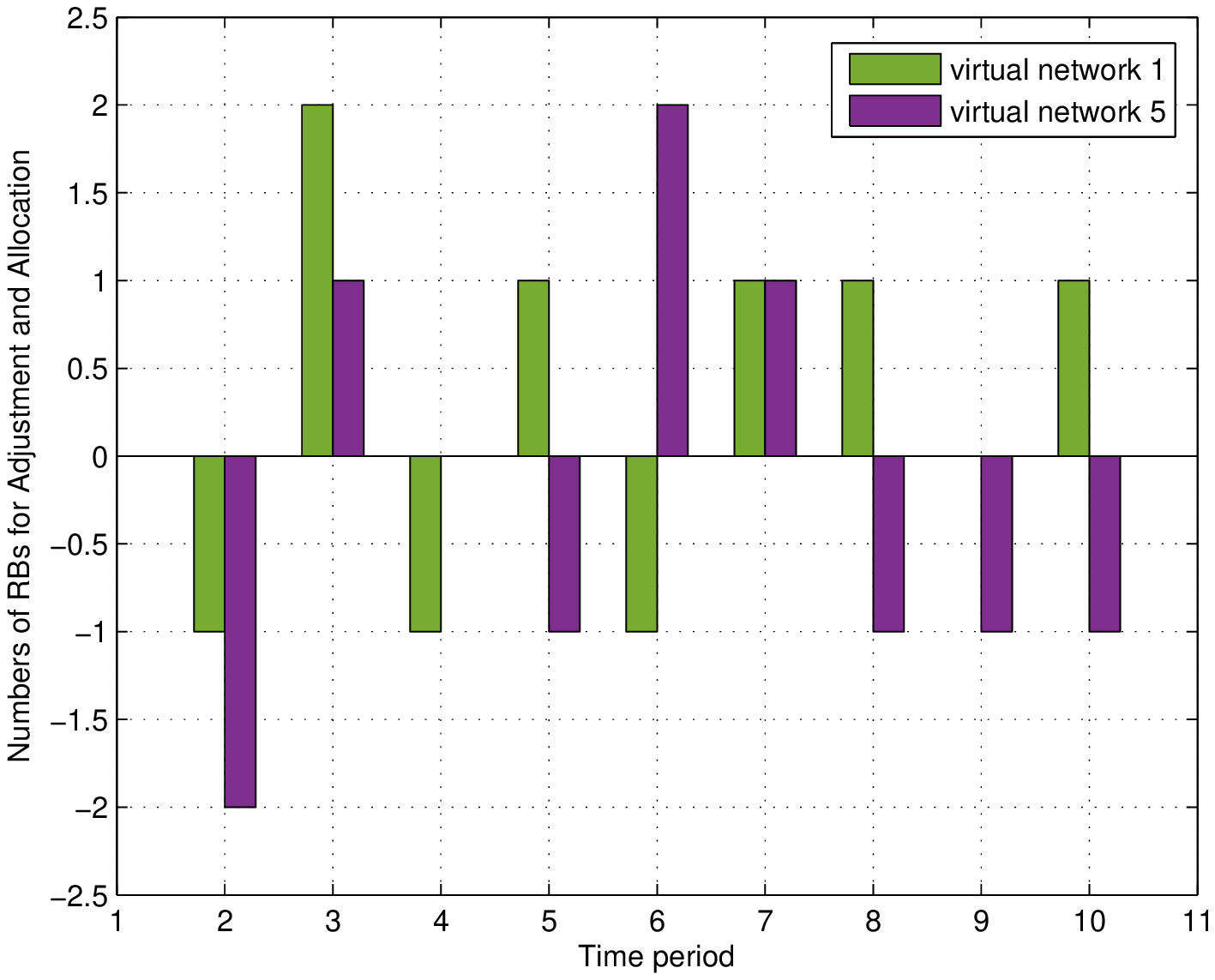}}
\caption{(a) The gap of ratio between the obtained and desired transmission rate in virtual networks $1$ and $5$. (b) The adjustment of the number of RBs in virtual networks $1$ and $5$.}
\label{fig:Feedback}
\end{figure*}

At first, we simulate the gap of ratio between the obtained and the desired transmission rate in two virtual networks. For the sake of comparison, the ratio of desired transmission rates in these two virtual network are set as 75\% and 25\%, respectively. Results in Fig. \ref{fig:Feedback}(a) reveal that in the first virtual network, the obtained transmission rate is always lower than  the desired rate. Hence, the RBs number adjustment and allocation in the highest class virtual network is necessary. Meanwhile, in the lowest class virtual network, the obtained transmission rate is in general higher than the desired rate. Hence, if the number of RBs can be adjusted and allocated dynamically at the end of each time period via the control loop by the SDN controller, the gap of ratio between the obtained and desired transmission rate will be decreased efficiently.\

Fig. \ref{fig:Feedback}(b) depicts the adjustment number of RBs in two virtual networks. Based on the gap of ratio between the obtained and desired transmission rate, the number of RBs will be adjusted and allocated dynamically by the SDN controller. The number of RBs should boost all the time to satisfy the desired transmission rate in the high class virtual network. By contrast, for the low class virtual network, although it also needs to increase RBs sometimes, the number of RBs will decrease after many time periods because the obtained transmission rate is higher than the desired transmission rate. Simulation results demonstrate that RBs need to increase or decrease momentarily to reduce the gap of ratio between the obtained and the desired transmission rate. In traditional M2M communication networks, the adjustment and allocation of RB number after each time period may be infeasible. However, it becomes possible in the proposed network architecture through the SDN controller and wireless network virtualization.\

\subsection{Convergence of the Proposed Algorithms}
This subsection investigates the convergence of the proposed scheme within different time periods. For simplicity, we also only focus on the highest class virtual network. The number of RBs is set as $5$ in each virtual network and the probability of false observation is $\epsilon=\varphi=0.1$. Other network parameters are the same as in subsection~\ref{sec:Simulation}-A. {\color{black}Specially, we have added the probability of preamble collisions in this part. In order to compare the performance, two probabilities of preamble collisions are considered, $Pr_s=0$ and $Pr_s=0.2$.}\

\begin{figure*}[!t]
\centering
\subfloat[]{\includegraphics[width=3in]{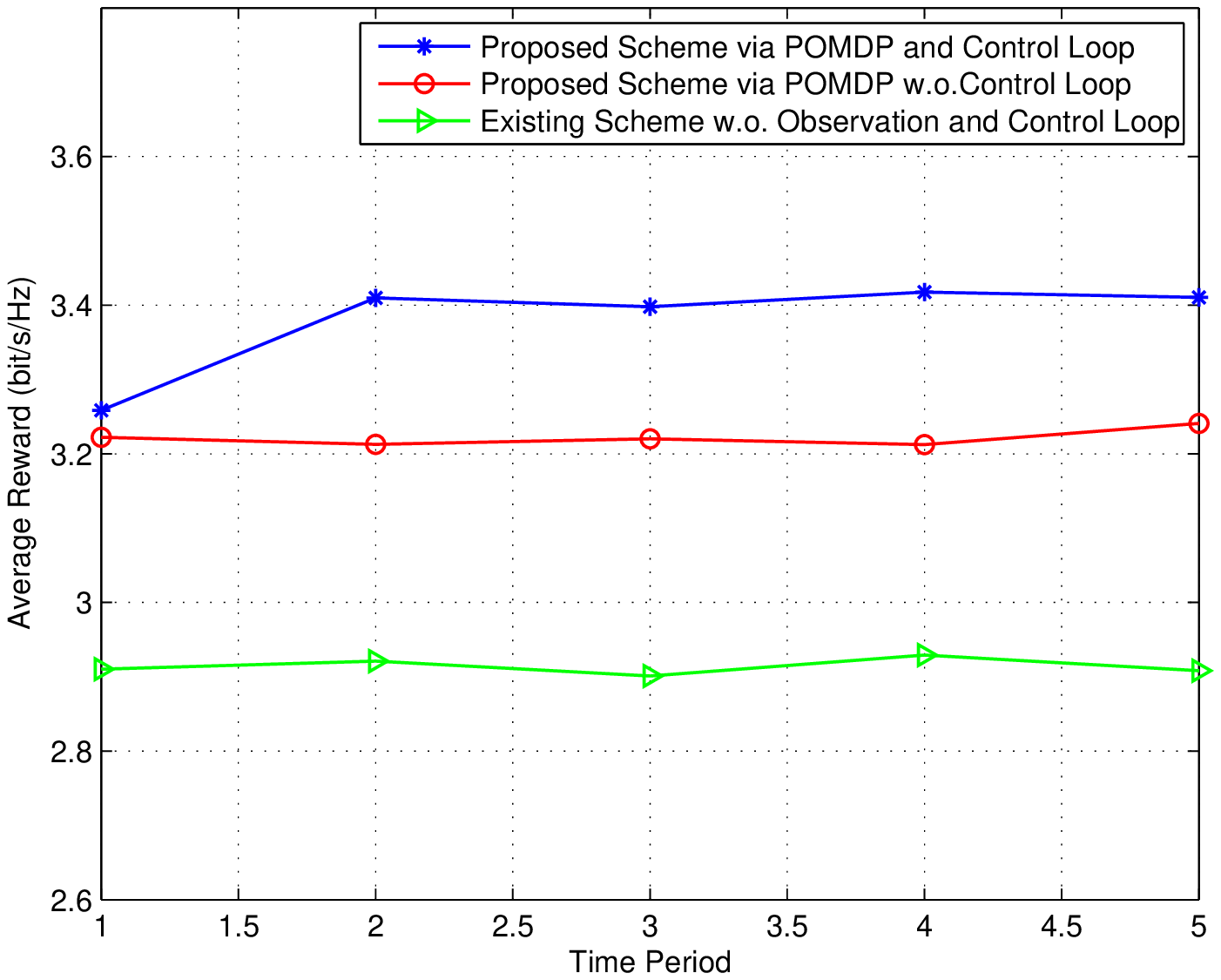}}
\subfloat[]{\includegraphics[width=3in]{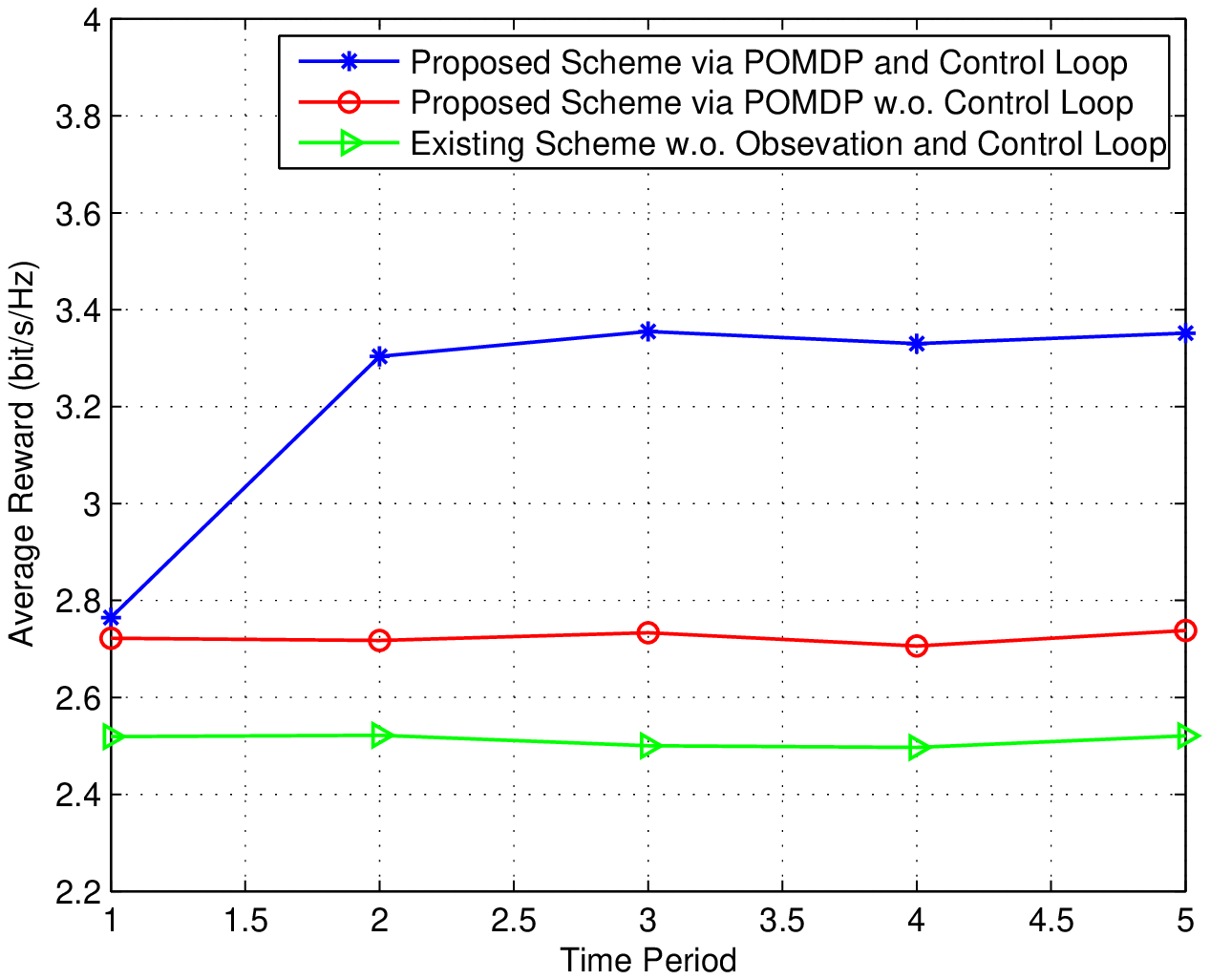}}
\hfil
\subfloat[]{\includegraphics[width=3in]{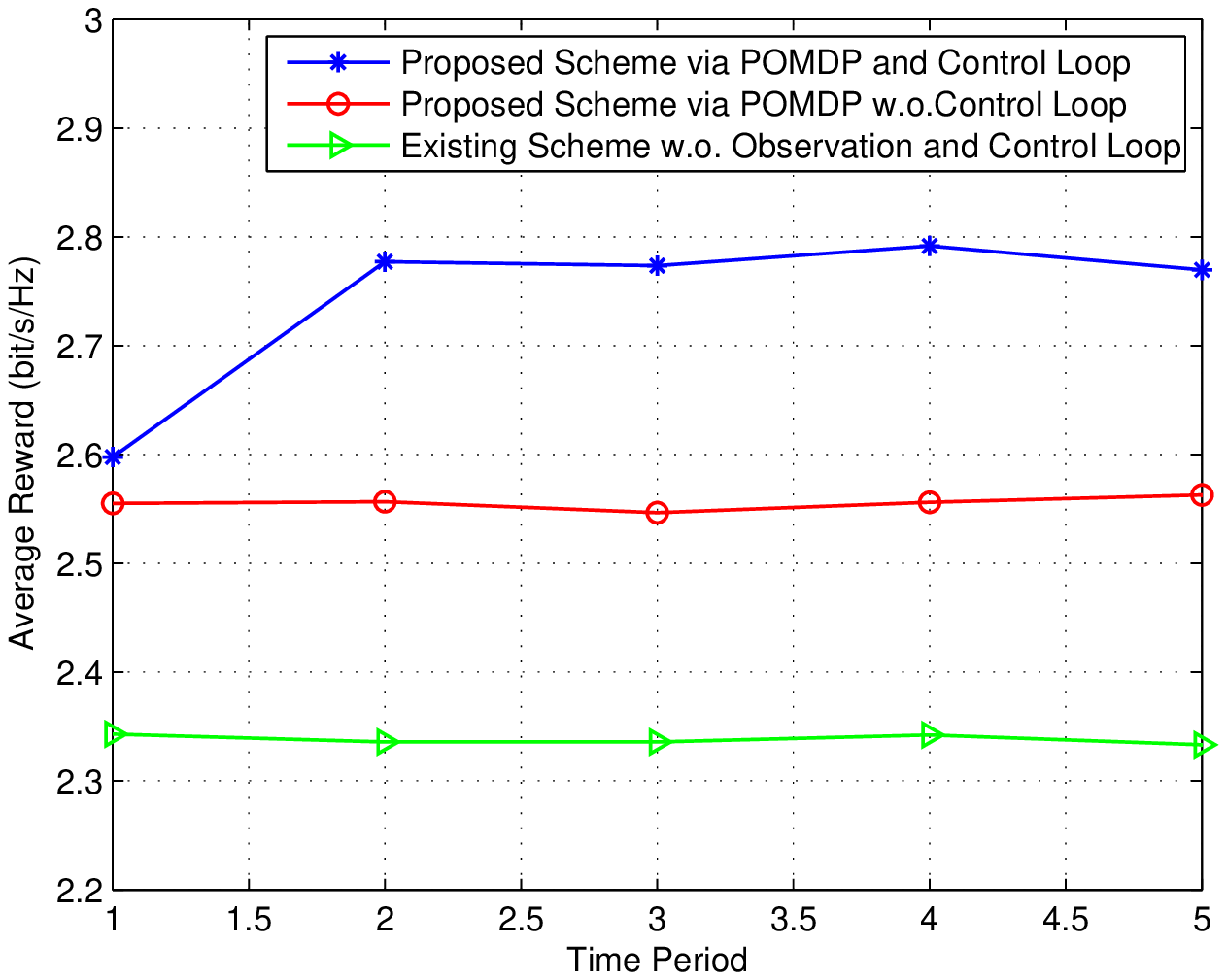}}
\subfloat[]{\includegraphics[width=3in]{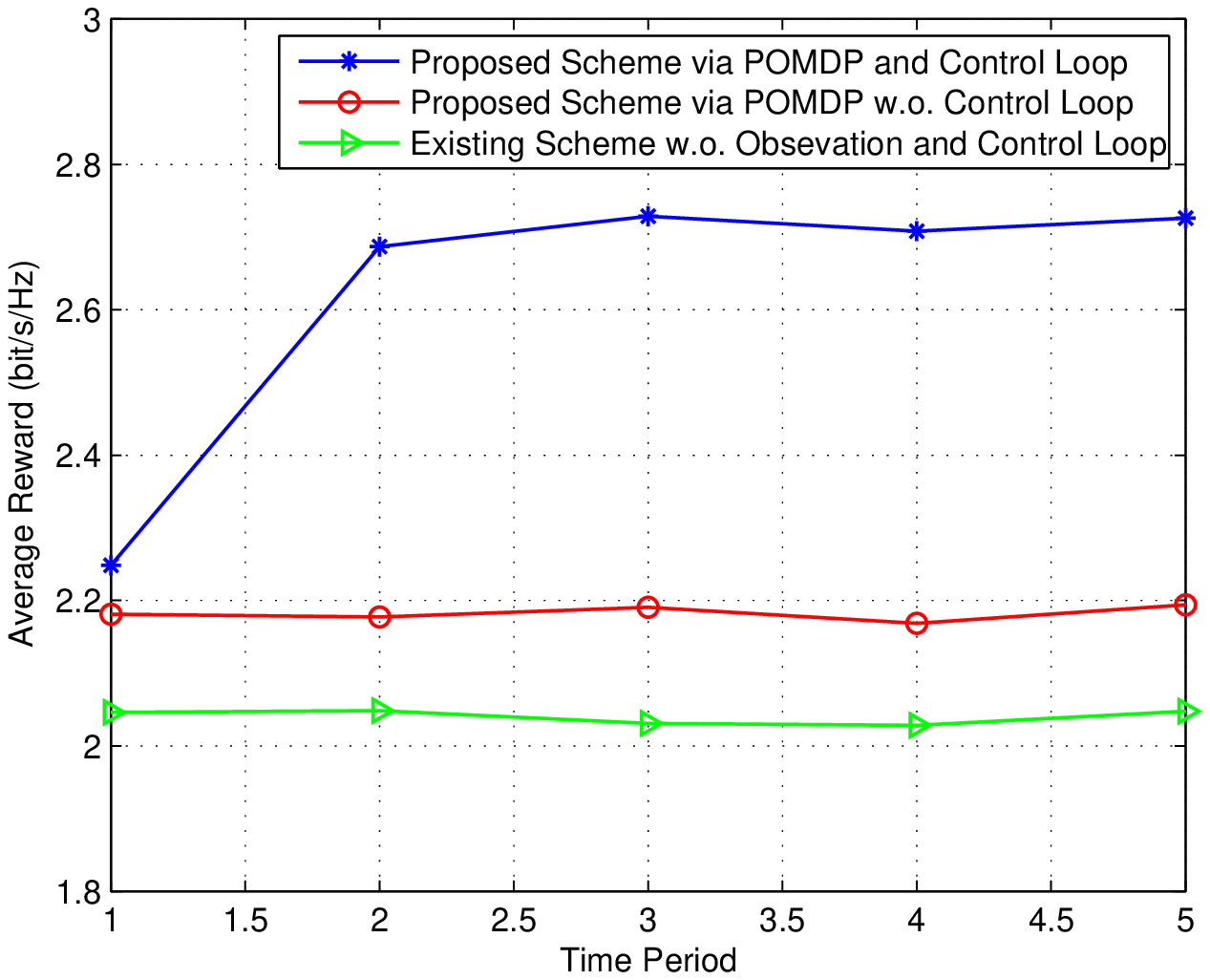}}
\caption{Average reward with different time periods in the (a) homogeneous traffic scenario ($Pr_s=0$), (b) heterogeneous traffic scenario ($Pr_s=0$), (c) homogeneous traffic scenario ($Pr_s=0.2$) and (d) heterogeneous traffic scenario ($Pr_s=0.2$).}
\label{fig:Rewardtime}
\end{figure*}

As shown in Fig. \ref{fig:Rewardtime}(a) and Fig. \ref{fig:Rewardtime}(b), the comparison of the proposed scheme without the control loop and the existing scheme without observation and the control loop when $Pr_s=0$, illustrate the average reward improvement by the proposed scheme. Obviously, it can be found that the proposed scheme via POMDP and the control loop has higher reward than other two schemes. Of course, the average reward by the proposed scheme via POMDP without the control loop is also larger than that the existing scheme without POMDP and the control loop. It can be explained that RBs with better performance may be selected through state observation. With POMDP optimization policy, the observation feature is highlighted by comparing it with existing schemes. {\color{black}Furthermore, since RBs can be adjusted and allocated continually and dynamically by the SND controller after each time period, then the ratio of the obtained transmission rate can be closed to the desired one. Therefore, the obtained transmission rate can keep a stable state to satisfy the desired one in each time period.}\

In addition, it should be noticed that there are also some differences  in two different traffic scenarios even though their tendency is consistent. Fig. \ref{fig:Rewardtime}(a) demonstrates the reward of the transmission rate with the proposed scheme in the homogeneous traffic scenario. It can be seen that the performance in the proposed scheme with the control loop outperforms the schemes without the control loop. Simulation results also reveal that the average reward in different schemes tends to be stable after the first time period. However, for the  homogeneous traffic scenario, the difference between the proposed scheme with and without the control loop is not prominent, and the gap is only about $0.2$ bit/s/Hz.\

On the contrary, for the heterogeneous traffic scenario, the advantage of the control loop can be evidenced distinctly. Fig. \ref{fig:Rewardtime}(b) demonstrates that the system performance can be improved significantly by the proposed scheme with the control loop. After the first time period, the number of RBs can be adjusted and allocated through the control loop. Due to the fierce competition for resources in the heterogeneous traffic scenario, more RBs will be allocated when the obtained transmission rate is lower than the desired transmission rate. It can be easily seen that the difference between the proposed scheme with and without the control loop is  reaching nearly $0.7$ bit/s/Hz. Similarly, with the variation of time period, the average reward tends to stable state for each scheme in the heterogeneous traffic scenario.\

{\color{black}Besides, as can be seen in Figs. \ref{fig:Rewardtime} (c) and (d), when the probability of preamble collisions is $0.2$, the average reward decreases obviously in both homogeneous and heterogeneous traffic scenarios, which compares with the situation of no preamble collisions. However, it can be found that the other results and tendency are all the same. It also demonstrates that the proposed scheme significantly outperforms the existing schemes even with preamble collisions.}\

\section{Conclusions and Future Work}\label{sec:Conclusion}
In this paper, we proposed a novel framework for M2M communications in software-defined cellular networks with wireless network virtualization. In the proposed framework, RBs, eNodeBs and MTCDs are virtualized as virtual resources. We formulated the random access process as a POMDP, by which MTCDs can select proper RBs to achieve the maximum transmission rate. In addition, a feedback and control loop was developed to adjust and allocate RBs by the SDN controller after each time period. With virtual resource allocation in each virtual M2M network, the obtained transmission rate approaches the desired transmission rate and the system performance can be improved through the control loop. Simulation results demonstrated that, with the proposed framework, the number of RBs can be dynamically adjusted according to the gap of ratio between the obtained and the desired transmission rate in each virtual network. Moreover, the transmission rate achieved by MTCDs can be improved significantly. Future work is in progress to consider   energy consumption and cooperative communications in our framework.\


\bibliographystyle{IEEEtran}
\bibliography{limengreference}

\end{document}